\documentstyle[aps,prb,amsfonts,amsmath,amssymb,eqsecnum,
                multicol,fancybox,rotate,epsfig]{revtex}

\def\beginmcols
{\begin{multicols}{2}%
\narrowtext%
}
\def\endmcols{%
\end{multicols}%
\widetext%
}

\begin{document}

\newcommand{\vc}[1]{\mathbf{#1}}

\title{Critical dynamics at incommensurate phase transitions and
\boldmath{$NMR$} relaxation experiments}

\author{B. A. Kaufmann, F. Schwabl, and U. C. T\"auber}

\address{Institut f\"ur Theoretische Physik, Physik-Department der Technischen
Universit\"at M\"unchen, \\ James-Franck-Stra\ss e, 85747 Garching, Germany}

\date{\today}

\maketitle

\begin{abstract}
We study the critical dynamics of crystals 
which undergo a second-order phase transition from a high-temperature  
normal phase to a structurally incommensurate ($IC$) modulated phase.
We give a comprehensive description of the critical dynamics of 
such systems, e.g. valid for crystals of the 
$\mbox{A}_{2}\mbox{B}\mbox{X}_4$ family.
Using an extended renormalization scheme, we present a framework 
in which we analyze the phases above and below the critical temperature
$T_I$.  
Above $T_I$, the crossover from the critical behavior to the mean-field
regime is studied. Specifically, the resulting width of the critical 
region is investigated.
In the $IC$ modulated phase, we consider explicitly 
the coupling of the order parameter modes to one-loop order. 
Here the Goldstone anomalies and their effect on measurable 
quantities are investigated. 
We show their relation with the postulated phason gap. 
While the theory can be applied to a variety of experiments,  
we concentrate on 
quadrupole-perturbed nuclear magnetic resonance ($NMR$) experiments.
We find excellent agreement with these dynamical measurements and 
provide answers for 
some questions that arose from recent results.

\end{abstract}


%
%

\beginmcols

\section{Introduction}
This paper is concerned with the critical dynamics of 
crystals undergoing a second-order phase transition
from a high-temperature normal ($N$) phase to a structurally incommensurate
($IC$) modulated phase.
In the $IC$ phase, the translational symmetry of the lattice is broken by a
modulation in such a way that the characteristic wave vector ${\vc q}_I$
is a non-rational multiple of a basic lattice vector.
The occurrence of incommensurate modulations is in general understood as a 
consequence of competing interactions. \cite{Sel92}
The most important characteristic of these systems
is that the ground state does not depend 
on the actual phase of the incommensurate modulation at a lattice site.
This implies that the initial phase of the modulation wave is arbitrary 
and one must take into account a phase shift degeneration of the ground
state energy.
Consequently, not only the amplitude of the modulating vector is 
required to characterize each configuration, but 
in addition the phase at a arbitrary lattice site must be fixed.
Therefore,  a two-dimensional order parameter ($OP$) has to be 
employed in order to describe the phase transition from a $N$ phase to an 
$IC$ modulated phase. \cite{Bru78a}
Interesting static properties, e.g.,
the very rich phase diagrams of systems with competing interactions,
emerge. \cite{Cum90}

However, in this work we concentrate on dynamical properties.
Considering fluctuations of the $OP$, 
the normal modes can be expressed in terms of the 
transverse and longitudinal components $\psi^\perp$ and $\psi^\parallel$  
in the two-dimensional $OP$ space. \cite{Bru78a}
The fluctuations of $\psi^\perp$ and $\psi^\parallel$ can be 
identified with the fluctuations of
the phase and the amplitude of the modulation in the crystal. \cite{Bru78b} 
As a consequence of the $OP$ being two-dimensional, 
the lattice dynamics of structurally incommensurate phases
shows some peculiar effects which are different from ordinary 
crystalline phases.
Namely, below the transition temperature $T_I$ 
two non-degenerate branches of modes appear in the dynamical
spectrum. \cite{Bru78b}
The ``amplitudon'' branch, connected with the fluctuations of the
amplitude of the incommensurate modulation, exhibits common soft-mode
behavior. 
In addition the ``phason'' branch represents the massless Goldstone
modes of the system, here originating of the invariance of the crystal energy 
with respect to a phase shift.
Because of the massless Goldstone modes \cite{Gol61,Wag66} present 
in the entire $IC$ phase, new types of anomalies may occur. 
Examples of such anomalies were discussed in the literature 
before. \cite{Nel76,Maz76,Sch92,Sch94}
Thus we expect some peculiar features of the dynamics in the $IC$ 
modulated phase stemming 
from the Goldstone modes and their coupling to the other $OP$ modes. 

The purpose of this paper on the one hand is to provide 
a general framework for the 
analysis of the critical dynamics above and below the $N$/$IC$ 
phase transition. 
The theoretical description of such systems is based on an $O(2)$
symmetric time-dependent Ginzburg-Landau model, with purely
relaxational behavior of the non-conserved order parameter. \cite{Bru78b} 
The more general $O (n)$-symmetric model 
has been widely studied above the
critical temperature by means of the dynamical renormalization group.
\cite{Hal72,Dom75} 
Below the critical temperature, the $O(n)$ symmetry
is spontaneously broken; 
as mentioned above parallel and perpendicular fluctuations have to be
distinguished.
We will start from the field-theoretical model of 
incommensurate phase transitions and derive the corresponding dynamical 
Janssen-De~Dominicis functional, \cite{Dom76,Jan76} 
which provides us with the framework to calculate 
interesting theoretical properties and correlations functions, which are
required for the interpretation of experimental data.

Furthermore, we intend to give a comprehensive theoretical 
description that goes beyond the mean-field or quasi-harmonic approach 
for the $IC$ phase and which is missing to date.
We present an explicit renormalization-group analysis to one-loop 
order above and below the critical temperature $T_I$.
The renormalization group theory will lead us beyond the 
mean-field picture and provide some new insight on the effects 
of the Goldstone modes on the dynamical properties below $T_I$.
Some specific features of the
Goldstone modes were discussed for the statics by Lawrie 
\cite{Law81} and for the dynamics in Refs. \onlinecite{Tae92} and
\onlinecite{Sch94}.
In the present paper, we extend the analysis of the 
$O(n)$-symmetric model, specifically
for the case $n=2$.
We also consider the crossover behavior from the 
classical critical exponents to the non-classical ones in detail, both 
above and below the critical temperature (see also Ref. 
\onlinecite{Tae93}).

Furthermore our model is employed to analyze specific experiments. 
Quadrupole-perturbed nuclear magnetic resonance ($NMR$) is an 
established method
to investigate $IC$ phases in a very accurate way. \cite{Bli86}
In this probe, the interaction of the nuclear quadrupole moment 
($Q$) of the
nucleus under investigation with the electric field gradient ($EFG$) at its
lattice site is measured. 
The fluctuations of the normal modes give rise to a fluctuating $EFG$, which is
related to the transition probabilities between the nuclear spin levels.
As a consequence, the relaxation rate $1/T_1$ of the spin-lattice relaxation is
given by the spectral density of the $EFG$ fluctuations at the Larmor
frequency.
We calculate the $NMR$ relaxation rate with our theoretical model, 
and compare our
findings with the experimental data.
Our results may be used to interpret a variety 
of experimental findings; however, here we 
will restrict ourselves to the analysis of $NMR$ experiments.
The theory presented here 
is appropriate for the universality class containing, e.g., 
the crystals of the $\mbox{A}_{2}\mbox{B}\mbox{X}_4$ family. 
Some very precise $NMR$ experiments on these crystals were 
performed over the past years. \cite{Bli86,Wal94}
Above $T_I$, these data can be used to analyze the critical
dynamics in a temperature range of $T-T_I = 100 K$ more closely.
Below $T_I$, an identification of relaxation rates, 
caused by fluctuations of the amplitude and the phase, respectively, at   
special points of the $NMR$ frequency distribution is possible.
Therefore the relaxation rates 
$1/T_A$ and $1/T_\phi$, referring to the 
critical dynamics of the two distinct excitations (``amplitudons" and
``phasons"), can be studied separately.
These experiments led to some additional open questions.
Above the critical temperature $T_I$, a large region was reported where
non-classical critical exponents were found. \cite{Hol95}
Below $T_I$, the presence of a phason gap is discussed in order to clarify some
experiments as well as the theoretical understanding. 
\cite{Top89,Hol95}
We will show how these questions can be resolved within the framework of
our theory.

This paper is organized as follows:
In the following section we introduce the model free energy for a system 
that reveals a $N$ to $IC$ phase transition.  
The dynamics of the amplitude and phase modes is described by
Langevin-type stochastic equations of motion, and
we give a brief outline of the general dynamical perturbation theory.
In Sec. III, the connection between the $NMR$ experiments and the
susceptibility calculated within our theory is discussed.
We shall see that the spectral density functions are closely related to
the measured relaxation times.
The high-temperature phase is analyzed in Sec. IV.
Above $T_I$, scaling arguments are used to derive the critical exponents
for the relaxation rate. The crossover from non-classical to
classical critical behavior is discussed by means of a renormalization group
analysis, and we comment on the width of the critical region.
In Sec. V,  we apply the renormalization group 
formalism to the low-temperature phase.
The susceptibility, containing the critical dynamical behavior for 
the amplitudon and phason modes, is calculated to one-loop order.
Specifically, the influence of the Goldstone mode is investigated.
In the final section we shall discuss our results and give some conclusions. 

\section{Model and Dynamical Perturbation Theory}
\subsection{Structurally incommensurate systems}
We want to study second-order phase transitions from a high-temperature
normal ($N$) phase to a structurally incommensurate ($IC$) modulated phase
at the critical temperature $T_I$.
The real-space displacement field corresponding to the one-dimensional
incommensurate modulation can be represented by its normal mode coordinates
$Q({\vc q})$. \cite{Bru78a}
We will treat systems with a star of soft modes \cite{La80} 
consisting only of two
wavevectors ${\vc q}_I$ and ${-\vc q}_I$ along one of the principal
directions of the Brillouin zone, e.g. substances of the 
$\mbox{A}_{2}\mbox{B}\mbox{X}_{4}$ family. \cite{Bru78a}
Because the incommensurate modulation wave is in most
cases, at least close to $T_I$, a single harmonic function of space,
the primary Fourier components $\langle Q({\vc q}) \rangle 
\propto \delta({\vc q} \pm {\vc q}_I)
e^{i \phi_0}$ with the incommensurate wavevectors $\pm {\vc q}_I$
are dominating.
Using $Q({\vc q})$ as a primary order parameter of the
normal-to-incommensurate phase transition in the Landau-Ginzburg-Wilson
free energy functional, diagonalization leads to \cite{Bru78a}
\begin{align}       
H[ \{ \psi_\circ^\alpha \} ] = &\frac{1}{2} \sum_{\alpha= \phi,A} \int_{\vc k}
(r_\circ + k^2) \psi_\circ^\alpha ({\vc k}) \psi_\circ^\alpha (-{\vc k})
\\
& +  \frac{ u_\circ }{4 !} \sum_{\alpha,\beta = \phi,A}
\int_{{\vc k}_1} \dots \int_{{\vc k}_4} \nonumber \\
& \times \psi_\circ^\alpha ({\vc k}_1) \psi_\circ^\alpha ({\vc k}_2) 
\psi_\circ^\beta ({\vc k}_3) \psi_\circ^\beta ({\vc k}_4) \;
\delta(\sum_{l=1}^{4} {\vc k}_l)  \ , \nonumber 
\end{align}
with new Fourier coordinates 
$\psi_\circ^\phi({\vc k})$ and $\psi_\circ^A({\vc
k})$ in the $OP$ space. Here, we have introduced the abbreviations
$\int_{k}^{} ... = \frac{1}{(2 \pi)^d} \int_{}^{} d^d k ... $, and
$\int_{\omega}^{} ... = \frac{1}{2 \pi} \int_{}^{} d \omega ... $ .

Below the phase transition, the 
fluctuations of $\psi_\circ^\phi$ and $\psi_\circ^A$ can be 
identified with the fluctuations of the phase and the 
amplitude of the displacement field, 
named {\it phason} and {\it amplitudon}. \cite{Bru78b} 
The wavevector ${\vc k}$ indicates the derivation from the incommensurate 
wavevector ${\vc q_I}$,
\begin{align}
{\vc k} = {\vc q} \mp {\vc q}_I \ .
\end{align}
The parameter $r_\circ$ is proportional to the distance from the mean-field 
critical temperature $T_{\circ I}$ 
\begin{align}
r_\circ \propto T-T_{\circ I}
\end{align}
and the positive coupling $u_\circ$ gives the strength of the isotropic
anharmonic term. Unrenormalized quantities are denoted by the suffix
$\circ$. 

The functional $H[{\psi_\circ^\alpha}]$ 
describes the statics of the normal-to-incommensurate
phase transition. 
It represents the $n$-component isotropic Heisenberg model;
in the case $n=2$, it is also referred to as the XY model.
For the sake of generality, the $n$-component order parameter case will be
considered in the theoretical treatment.

\subsection{Critical Dynamics} 
The critical dynamics of the system under consideration is characterized
by a set of generalized Langevin equations for the ``slow" variables, which
in our case consist of the non-conserved order parameter fluctuations 
(because of the critical slowing-down in the vicinity 
of a phase transition). \cite{Hoh77}
The purely relaxational behavior \cite{Zey82} is described by the following
Langevin-type equation of motion
\begin{align}
\label{Lang}    
\frac{\partial}{\partial t} \, \psi_\circ^\alpha({\vc k},t)  =
- \lambda_\circ \, \frac{\delta H[\{
\psi_\circ^\alpha \}] }{\delta \psi_\circ^\alpha (-{\vc k}, t)} + 
\zeta^\alpha({\vc k},t) \ .
\end{align}
The damping is caused by the fast degrees of freedom, which are subsumed in 
the fluctuating forces $\zeta^\alpha$. 
According to the classification by Halperin and Hohenberg,
we are considering model A.  \cite{Hoh77}

The probability distribution for the stochastic forces 
$\zeta^\alpha$ is assumed to be Gaussian. Therefore 
\begin{align}
\langle \zeta^\alpha({\vc k},t) \rangle & = 0,  \\
\langle \zeta^\alpha({\vc k},t) \, \zeta^\beta({\vc k'},t')
\rangle  & = 2 \, \lambda_\circ \, \delta({\vc k} - {\vc k'}) \,
\delta(t - t') \, \delta^{\alpha \beta} \ ,
\label{stoch}
\end{align}
where the Einstein relation (\ref{stoch}) guarantees that the
equilibrium probability density is given by 
\begin{equation} \label{opvert}
P [ \{\psi_\circ^\alpha \} ] = 
\frac{e^{-H[ \{ \psi_\circ^\alpha \} ] }}{ 
\int {\cal D} [ \{ \psi_\circ^\alpha \} ] e^{-H[ \{ \psi_\circ^\alpha \} ] }} \quad.
\end{equation}

Following the dynamical perturbation theory 
developed by Janssen \cite{Jan76}
and De Dominicis, \cite{Dom76} we want to calculate the dynamical
properties of our system, e.g. the dynamical correlation functions.

First, the stochastic forces are eliminated, using equation 
(\ref{Lang}) and the
Gaussian distribution for the stochastic forces $\zeta^\alpha$. 
After a Gaussian transformation and the
introduction of auxiliary  Martin-Siggia-Rose \cite{Mar73} 
fields ${\tilde \psi}_\circ^\alpha $, the non-linearities occuring in the
initial functional are reduced.
A perturbation theory analogous to the static theory can now be implemented
on the basis of the path-integral formulation.
We define the generating functional
\begin{align}
Z[\{ {\tilde h}^\alpha \} &, \{ h^\alpha \}] \propto  \int {\cal
D}[\{ i {\tilde \psi}_\circ^\alpha \}] \, {\cal D}[\{ \psi_\circ^\alpha
\}] \, \nonumber \\
& \times e^{J[\{ {\tilde \psi}_\circ^\alpha \} , 
\{ \psi_\circ^\alpha \}] +
\int \! d^dx \! \int \! dt \sum_\alpha ({\tilde h}^\alpha \,
{\tilde \psi}_\circ^\alpha + h^\alpha \, \psi_\circ^\alpha)} \ , 
\end{align}
where the resulting Janssen-De Dominicis functional $J=J_0+J_{int}$ is split into
the harmonic part $J_0$ and the interaction part $J_{int}$,
\begin{align}   
\label{dyn_func_harm} 
J_0 &  [\{ {\tilde \psi}_\circ^\alpha \} , \{ \psi_\circ^\alpha\}] 
= \int_k \int_\omega \sum_\alpha \biggl[ \lambda_\circ \, 
{\tilde \psi}_\circ^\alpha({\vc k},\omega) \, {\tilde
\psi}_\circ^\alpha(- {\vc k},-\omega)  \nonumber \\
& - {\tilde \psi}_\circ^\alpha({\vc k},\omega) \, \Bigl[ i \omega +
\lambda_\circ \,  (r_\circ + k^2) \Bigr] \, \psi_\circ^\alpha(- {\vc
k},- \omega) \biggr] \ , \\
J_{int} & [\{ {\tilde \psi}_\circ^\alpha \} , \{
\psi_\circ^\alpha \}] 
= \frac{-\lambda_\circ \, u_\circ}{6} \, \int_{q_i} 
\int_{\omega_i} \,
\delta  ( \sum_i {\vc k}_i ) \, \delta  ( \sum_i \omega_i ) \nonumber  \\
 & \times \, \sum_{\alpha \beta} {\tilde \psi}_\circ^\alpha({\vc
k}_1,\omega_1) \, \psi_\circ^\alpha({\vc k}_2,\omega_2) \,
\psi_\circ^\beta({\vc k}_3,\omega_3) \, 
\psi_\circ^\beta({\vc k}_4,\omega_4)  \ .
\label{dyn_func}
\end{align}
The N-point Green functions 
$G_{\circ \, {\tilde \psi}^\alpha_i \psi^\alpha_j}({\vc k},\omega)$ 
and cumulants $G^c$
can be derived by appropriate derivatives of $Z$ and $\ln Z$ 
with respect to the sources ${\tilde h}^\alpha$ and $ h^\alpha$ .
Thus the standard scheme of perturbation theory can be applied.
Further details can be found in textbooks (Refs. \onlinecite{Ami84,Zin93}) 
and in Refs. \onlinecite{Jan76,Tae92}.

In addition we want to list some important relations that will 
be useful for the discussion.
The dynamical susceptibility gives meaning to the auxiliary fields by 
noting that it can be represented as a correlation function 
between an auxiliary field and the order parameter field \cite{Jan76}
\begin{align}
\chi_\circ^{\alpha \beta}({\vc x},t;{\vc x'},t') 
& = {\delta \langle \psi_\circ^\alpha({\vc x},t) \rangle \over \delta  
{\tilde h}^\beta({\vc x'},t')} \bigg \vert_{{\tilde h}^\beta =0} 
\nonumber \\
& = \langle \psi_\circ^\alpha({\vc x},t) \, \lambda_\circ \, {\tilde 
\psi}_\circ^\beta({\vc x'},t') \rangle \ .
\end{align}
Its Fourier transform
\begin{align}
\chi_\circ^{\alpha \beta}({\vc k},\omega) = \lambda_\circ \,  
G_{\circ \, {\tilde \psi}^\alpha \psi^\beta}({\vc k},\omega) \ .
\label{sus_cor1}
\end{align}
is associated with the Green functions $G_{\circ \, {\tilde \psi}^\alpha
\psi^\beta}$. 
The fluctuation-dissipation
theorem relates the correlation function 
of the order parameter fields and the imaginary part of the response function
\cite{Jan76} 
\begin{align}
\label{fluk_diss}
G_{\circ \, \psi^\alpha \psi^\beta}({\vc k},\omega) = 
2 \frac{\Im \chi_\circ^{\alpha \beta}({\vc k},\omega)}{\omega} \ ,
\end{align}
which will enter the calculation of the $NMR$ relaxation rate.
E.g., considering only the harmonic part $J_0$ of the dynamical functional and
carrying out the functional integration gives 
\begin{align}
G_{\circ \, \psi^\alpha \psi^\beta}({\vc k},\omega) =& 
\delta^{\alpha  \beta}
\frac{2 \lambda_\circ}{[\lambda_\circ(r_\circ+k^2)]^2 + \omega^2} \ .  
\label{G_harm}
\end{align}

Finally we want to introduce the vertex functions $\Gamma_{\circ \, {\tilde
\psi}^\alpha \psi^\beta}$, which are related to the
cumulants through a Legendre transformation. 
For example \cite{Jan76,Tae92}
\begin{align}
G_{\circ \, {\tilde \psi}^\alpha \psi^\alpha}^c ({\vc k},\omega) =
\frac{1}{\Gamma_{\circ \, {\tilde \psi}^\alpha \psi^\alpha}(-{\vc k},-\omega)} 
\ .
\label{cor_cum1}
\end{align}
The vertex functions are entering the explicit
calculation of $Z$ factors and the susceptibility 
in the renormalization group theory. 
The advantage of working with vertex functions is that they are represented by
one-particle irreducible Feynman diagrams only.

\section{$NMR$-Experiments and Spin-Lattice Relaxation}
\label{sec_nmr}
Quadrupolar perturbed nuclear magnetic resonance ($NMR$) can be
used to study the dynamics of phase transitions from a $N$
to a $IC$ modulated phase. \cite{Bli86}
In this method the interaction between the nuclear quadrupole moment
$Q$ and the electric-field gradient
($EFG$) $V$ is the dominant perturbation ${\cal H}_Q$ of the Zeeman 
Hamiltonian. 
Thus in the corresponding Hamiltonian
\begin{align}    
{\cal H} &= {\cal H}_{Z} + {\cal H}_{Q} 
\end{align}
next to the dominating Zeeman term ${\cal H}_{Z}$ one has to consider 
the quadrupole
interaction ${\cal H}_{Q}=\frac{1}{6} \sum_{j,k}^{} Q_{jk} V_{jk}$ as a
perturbation.  
The quadrupole moment operator $Q_{jk}$ is coupled linearly to the $EFG$
tensor  $V_{jk}$ at the lattice site. \cite{Bon70,Zum81}  \\
The fluctuations of 
$V_{jk}$ can be expressed via order parameter fluctuations, because
of the dominant linear coupling of the $EFG$ to the order parameter
\cite{Per87a,Per87b}
\begin{align}  
\delta V_{ij}({\vc x}, t) = A_{1ij} [\delta \psi^A(t) + i
\delta \psi^\phi(t)] e^{ i {\vc k} {\vc x} + \Phi_0} + c.c. \ .
\end{align}

We now briefly sketch how the $OP$ fluctuations determine the
relaxation rate.
The spin-lattice relaxation describes the return of the nuclear spin
magnetization $M$ in the direction of the external field
back to its thermal equilibrium value
following a radio frequency pulse. \cite{Abr86}
During that time the energy of the spin system is transferred to single
modes of the lattice fluctuations. \cite{Wal94}
Because the $EFG$ fluctuations can be written as $OP$ fluctuations, the
spin-lattice relaxation is determined by the spectral density 
functions of the
local $OP$ fluctuations at the Larmor frequency $\omega=\omega_L$.
The transition probabilities for nuclei with spin $I=\frac{3}{2}$ in three
dimensions are given by \cite{Abr86,Coh57}
\begin{align}   
\frac{1}{T_1} &= 
W \left( \pm \frac{3}{2}  \leftrightarrow \pm \frac{1}{2} \right)  
= \frac{\pi^2}{3}
\left[ J(V_{xy}, \omega_L) + J(V_{yz}, \omega_L) \right] \nonumber \\
& \propto \int_{BZ}^{}
\frac{\Im \chi_\circ^{\alpha \beta} ({\vc k},\omega_{L})}{\omega_{L}} 
= \int_{0}^{\Lambda} \frac{1}{2} \ k^2 \ G_{\circ \, \psi^\alpha \psi^\beta} 
({\vc k},\omega_L) dk  \ ,
\label{relax-time}
\end{align}
with the spectral density of the $EFG$ fluctuations
\begin{align}
J(V_{ij}, \omega)   =  
\int_{- \infty}^{\infty} \overline{V_{ij}(t) {V_{ij}}^{*}(t+\tau)} 
              e^{-i \omega \tau} d\tau  \ .
\end{align}
Measuring the spin-lattice relaxation thus yields information on
the susceptibility of the local fluctuations of the order parameter.

The spin-lattice relaxation was studied in great detail by means of
echo pulse methods both below and above $T_I$. \cite{Bli86,Wal94}
Below the critical temperature $T_I$, it is possible for the prototypic system
Rb$_2$ZnCl$_4$ to identify the relaxation rates $1/T^A_1$
and $1/T^\phi_1$, dominated in the plane-wave limit by 
the amplitudon and phason fluctuations,  respectively. \cite{Wal94}
Therefore, the dynamical properties of the order parameter fluctuations 
can be studied below the phase transition as well, and separately
for the two distinct excitations.

\section{High-temperature phase}
In this section, the critical behavior of the incommensurate 
phase transition above $T_I$ will be investigated.
On the basis of scaling arguments and the use of critical exponents, 
calculated within
the renormalization-group theory for the XY model in three dimensions and
model A, the temperature dependence of the $NMR$ relaxation rate is 
analyzed in the first subsection. 
Next we study the crossover scenario of the temperature 
dependence of the 
relaxation rate in the second subsection by means of the 
renormalization group theory.
Comparison with experimental data is made, and we comment on the width of
the critical region.

\subsection{Scaling laws for the relaxation time}
Above the phase transition, the thermodynamical average 
of the order parameter components is zero.
Because the structure of the correlation function of the order parameter 
does not change dramatically above $T_I$ (see section \ref{tgtc_renorm})
and the calculation of the relaxation 
rate $1/T_1$ involves the integration of the correlation function over all 
wavevectors,   
we will derive a form of the correlation function using scaling arguments.
Thus we are able to discuss the universal features of the relaxation rate
behavior when approaching $T_I$ from above. 
In the harmonic approximation we immediately get the correlation
function, which turns out to be the propagator of
our functional $J_0$,  [see Eq. (\ref{G_harm})]
\begin{align}
\langle \psi_\circ^\alpha({\vc k},t) \,  
\psi_\circ^\beta({\vc k'},t') \rangle &= \delta({\vc k} + {\vc k'}) 
\delta(\omega + \omega') \delta^{\alpha \beta} G_\circ (k,\omega) \ ,\\
G_\circ (k,\omega) &= \frac{2 \lambda_\circ}{[\lambda_\circ(r_\circ + k^2)]^2 +
\omega^2} \ .
\end{align}
The suffix $\circ$ will be omitted in the following discussion of this
subsection, because no renormalization will be considered here.

We want to exploit our knowledge about the critical region. The static scaling
hypothesis for the static response function states
\begin{align}
\chi(k) = A_\chi \cdot \hat \chi (x) \cdot k^{-2+\eta} \  
\end{align}
with the scaling function $\hat \chi$, a constant prefactor $A_\chi$,
the scaling variable $x = (k \xi)^{-1}$ ($\xi$ denoting the correlation
length), and the critical exponent $\eta$. 
Neglecting a frequency dependence for the
kinetic coefficient (Lorentzian approximation), 
the dynamic scaling hypothesis for the characteristic frequency of the $OP$
dynamics states
\begin{align}
\omega_{\varphi}(k)  &\equiv \lambda (k)/ \chi(k) \sim k^{z} \ ,
\label{dy_sc_hy}
\end{align}
and we can deduce
\begin{align}
\lambda (k)  &= A_\lambda \cdot \hat \lambda (x) \cdot 
k^{ z-2 + \eta} \ ,
\end{align}
with $ \hat \lambda (x)$ being the scaling function for the kinetic
coefficient and $A_\lambda$ a constant prefactor.
Notice that for fixed wavevector $k$ Eq. \ref{dy_sc_hy} leads to \cite{Hoh77}
\begin{align}
\omega_{\varphi}(k)  \sim \xi^{-z} \ . 
\end{align}
The correlation function $G(k,\omega)$ can now be rewritten in scaling form 
\begin{align}   
G(k,\omega)
= \Lambda \cdot \frac{1}{k^{z +2 - \eta}} \cdot \hat{f}(\hat{\omega},x) \ ,
\end{align}
with
\begin{align}
\Lambda &= \frac{2 \ A_\chi^{2}}{A_\lambda},  \ \ \hat{\omega} = 
\frac{A_\chi}{A_\lambda} \frac{\omega}{k^{z}}, \ \
x = \frac{1}{k \xi},  \nonumber \\
\hat{f}(\hat{\omega},x) &= \hat \chi(x) \cdot 
\frac{\hat \lambda(x)/{\hat \chi(x)}}{[\hat \lambda(x)/{\hat \chi(x)}]^2 +
\hat{\omega}^2}  \ ,
\end{align}
where the Lorentzian line shape is retained. Above $T_I$, this not a very crucial 
approximation, because the shape of the correlation function does 
not change in a first-order renormalization group analysis, 
as we will see in the next section.
To calculate the relaxation rate $1/T_1$, one has to evaluate the integral
[see Eq. (\ref{relax-time})]
\begin{align}
\frac{1}{T_1} 
& \propto  \int_{0}^{\Lambda} \frac{1}{2} \ k^2 \ G(k,\omega_L) dk \\
& \propto \Lambda \cdot \int_{BZ}^{}k^2 dk \ k^{-z -2 + \eta} 
\hat{f}(k \omega_L^{-1/z}, k \xi) \ . \nonumber 
\end{align}
With $u = k \omega_L^{-1/z}$ and $v= k \xi$
we introduce new variables 
\begin{align}
\label{varrho}
\varrho &= \sqrt{u^2+v^2} = k \sqrt{\omega_L^{-2/z} + \xi^2}  \\
\intertext{and}  
\label{varphi}
\tan \varphi & = \frac{v}{u} = \frac{\xi}{\omega_L^{-1/z}} \ .
\end{align}
This leads to the relation
\begin{align}
\frac{1}{T_1} 
= \Lambda \cdot \left( \sqrt{\omega_L^{-2/z} + \xi^2} \right)^{z -1 -\eta} \ 
I_{\varrho} \left( \hat f(\varrho,\varphi) \right) \ ,
\end{align}
where the integral $I_{\varrho}$ does not contribute to the leading temperature
dependence. 

The temperature dependence of the relaxation rate can now easily be found 
in the limits where the Larmor frequency or the 
frequency, of the critical fluctuations, respectively, dominate the integral 
and its prefactor.

\subsubsection{Fast-motion limit ($\omega_{L}/\omega_{\varphi} \ll 1 $)}
For temperatures very far above the 
critical temperature $T_I$, the characteristic frequency
is larger than the Larmor frequency. Thus the temperature dependence
of the $OP$ fluctuations determines the temperature dependence of the
relaxation rate; the value of the Larmor frequency should not play any
role. \cite{Rig84} For the integral
\begin{align}
\label{results_fml}
\frac{1}{T_1}  &=
\Lambda \cdot 
\left( \omega_{L}^{-2/z} \left[ 1 + \left( 
\omega_{L}/ \xi^{-z} \right)^{2/z}
\right] \right)^{(z - 1 - \eta)/2}
\cdot I_{\varrho} \nonumber \\
\end{align}
we obtain with $\tan \varphi =$ const. [see Eq. \ref{varphi}]
\begin{align}
\frac{1}{T_1} &\propto \xi^{z-1 - \eta} 
 = \left( \frac{T-T_{I}}{T} \right)^{-\nu \cdot (z-1-\eta) } 
 \ .
\end{align}
Taking the values for the critical exponents from Table \ref{tab1}, we find 
\begin{align}
\frac{1}{T_1} \propto \ \left( \frac{T-T_I}{T}
\right)^{-0.663} \ \ .  
\end{align}

This can be compared with the experimental results for 
Rb$_2$ZnCl$_4$ by Holzer et al. \cite{Hol95}, 
who found for the leading scaling
behavior of the relaxation rate, following the 
temperature-independent region, the exponent $-0.625$.
 
\subsubsection{Slow-motion limit ($\omega_{L}/ \omega_{\varphi} \gg 1$)} 
In the vicinity of the critical temperature $T_I$ critical slowing down
will occur. This means that the characteristic frequency $ \omega_{\varphi}$
is approaching zero and will fall below the value of the Larmor frequency.
\cite{Rig84} Thus, the characteristic
time scale of the $OP$ fluctuations is slower than the
experimental time scale. For the temperature dependence of the 
relaxation rate
\begin{align}  
\frac{1}{T_1}
& \propto 
\Lambda \cdot 
\left( \xi^{2} \left[ \left( 
\xi^{-z}/\omega_{L} \right)^{2/z} + 1
\right] \right)^{(z - 1- \eta)/2}
\cdot I_{\varrho} \ , \nonumber 
\end{align}
we now obtain
\begin{align} 
\frac{1}{T_1} &\propto \omega_{L}^{-(z-1 -\eta)/z} \cdot \text{const} \ .
\end{align}
Taking the values for the critical exponents from Table \ref{tab1} again 
\begin{align}
\frac{1}{T_1} \propto \ \omega_{L}^{-0.49} \ .
\end{align}

This is in good agreement with the experimental result \cite{Hol95} for
Rb$_2$ZnCl$_4$ that the value of
the relaxation rate near $T_I$ scales as $\omega_L^{-0.5}$ 
for different Larmor frequencies. 

We want to stress that the transition from the fast- to the slow-motion limit
is a property of the integral entering the calculation of the
relaxation rate. 
Because the susceptibility is evaluated at fixed $\omega_L$, there exists
for  a lower boundary for the integral $T \approx T_I$. 
This means that the
transition from the temperature-dependent to the temperature independent
behavior near $T_I$ is fixed by the scale $\omega_L$.
It should also be mentioned that our results reproduce those
obtained earlier by Holzer et al. (see Ref. \onlinecite{Hol95}), 
when the van Hove approximation for $z$ is used ($z \approx 2$).

\subsection{Renormalization-group analysis above $T_I$}
\label{tgtc_renorm}
Our investigations concerning the critical behavior above $T_I$ in the last
section led to fair agreement with experimental data.
It was possible to gain the critical exponents for the 
frequency and temperature dependence of the relaxation rate in a 
quantitatively accurate way.
Furthermore, we obtained a qualitative understanding of 
the transition of the relaxation rate $1/T_1$ from
the slow- to the fast-motion limit.
This transition is caused by the characteristic frequency of the
order-parameter dynamics $\omega_\varphi$ approaching zero, i.e., the
critical slowing down near $T_I$. 
This renders the Larmor frequency
$\omega_L$ the dominating time scale, $1/T_1$ becomes temperature-independent 
near $T_I$.

At this point we want to consider 
what happens upon leaving the region near the transition
temperature and going to higher temperatures.
Our results (\ref{results_fml}) for the fast-motion 
limit are based on the assumption
that fluctuations are very important.
We used non-classical exponents and scaling arguments valid in the critical
region.
The question arises how large this region of temperature, where
the system displays the non-classical behavior calculated in the last section,
will be.
Increasing temperature diminishes the effect of fluctuations.
One would expect that at some temperature Gaussian behavior 
should emerge.
We shall apply the renormalization group theory to one-loop 
approximation in order to describe the transition from the 
fluctuation-dominated behavior near $T_I$ to a temperature region
where the mean-field description should be valid.
It is obvious that the properties of the integral, responsible for the transition
between the slow and the fast motion limit, will account for the crossover
of the leading scaling behavior when the temperature is increased. 
In order to discuss the crossover within this analysis a modified minimal-subtraction
prescription is employed. 
This scheme was first introduced by Amit and
Goldschmidt \cite{Ami78} and subsequently 
explored by Lawrie for the study of Goldstone
singularities. \cite{Law81}
It can comprise exact statements in a certain limit. 
Below $T_I$, this is the regime
dominated by the Goldstone modes alone; in the region above $T_I$, which we will
consider in this section, the mean-field result is used. In this
scheme, in  addition the standard field-theoretical formulation 
of the renormalization group is neatly reproduced.
Following the arguments of Schloms and Dohm \cite{Sch89} and 
Ref. \onlinecite{Tae92} we can 
refrain from the $\varepsilon$ expansion, with
\begin{align}
\varepsilon = 4-d
\end{align}
defining the difference from the upper critical dimension of the $\phi^4$
model. 
This is motivated by the following. 
Above $T_I$ the Gaussian or zero-loop theory
becomes exact in the high-temperature limit.
The critical fixed point, i.e., the Heisenberg fixed point,
dominating the behavior of the system near the critical temperature is
calculated to one-loop order.
The main interest here lies in the crossover behavior between these 
two fixed points, which is calculated to one-loop order, too.
Thus no further approximations are necessary to be consistent.
 
Very close to the critical temperature 
$T_I$ an $\varepsilon$-expansion or Borel resummation
\cite{Sch89} would of course be inevitable in order
to obtain better values for the critical exponents.
A description of the generalized minimal subtraction scheme is for example
given in Refs. \onlinecite{Ami78,Law81,Tae92} and \onlinecite{Fre94}.
A crossover into an asymptotically Gaussian theory is described by this method
in Ref. \onlinecite{Tae93}.

\subsubsection{Flow equations}
Our aim is to calculate the wavenumber and frequency dependence of
the susceptibility to one-loop order.  
The field renormalization is zero to one loop order. Thus  
we will not take into account corrections to the static exponent
$\eta$ and corrections to the mean-field value of the dynamic exponent 
$z \approx 2$. This leaves $r_\circ$ and $u_\circ$ as the only 
quantities to be renormalized. \cite{Tae93}

There is a shift of the critical temperature from the mean-field result 
$T_{\circ I}$ to the ``true" transition temperature $T_I$. 
In order to take this shift into account, 
a transformation to a new temperature variable, 
being zero at the critical temperature $T_I$, is performed.
This new variable will be denoted again as $\tau_\circ$.  
The only renormalized quantities are then written as 
\begin{align}
\label{tau_HT}
\tau &= Z_r^{-1} \tau_\circ \mu^{-2} \\
\label{u_HT}
u &= Z_u^{-1} u_\circ A_d \mu^{-\varepsilon} \ .
\end{align}
Here, the geometric factor $A_d$ is chosen \cite{Sch89} as
\begin{align}
A_d  = \frac{\Gamma(3-d/2)}{2^{d-2} \pi^{d/2}(d-2)} \ .
\end{align}

For the non-trivial $Z$ factors one finds in the generalized minimal
subtraction procedure (see App. \ref{app1})
\begin{align}
\label{Z-factors}
Z_u &= 1 + \frac{n+8}{6 \varepsilon} u_\circ A_d \mu^{- \varepsilon} 
\frac{1}{(1+\tau_\circ/\mu^2)^{\varepsilon/2}} \ , \\
Z_r &= 1 + \frac{n+2}{6 \varepsilon} u_\circ A_d \mu^{- \varepsilon} 
\frac{1}{(1+\tau_\circ/\mu^2)^{\varepsilon/2}} \ .
\label{Z_gt}
\end{align}
Setting $\tau_\circ=0$ the familiar renormalization constants for the
$n$-component $\phi^4$ model are recovered.
In general here, however, 
the $Z$ factors are functions of both $u_\circ$ and $\tau_\circ$. \cite{Ami78}
In the next step the fact that the
unrenormalized $N$-point functions do not depend on the scale $\mu$ is 
exploited and the
Callan-Symanzik equations are derived. \cite{Ami84} 
The idea behind that is to connect
via the renormalization-group equations the uncritical theory, which can be
treated perturbationally, with the critical theory displaying infrared
divergences. 
The resulting partial differential equations can be solved with the method of
characteristics ($\mu(l)=\mu \, l$). 
With the  definition of Wilson's flow functions 
\begin{align}
\zeta_\tau (l) &= \left. \mu \frac{\partial}{\partial \mu} \right|_0 
\ln \frac{\tau}{\tau_\circ} \ , \\
\beta_u (l) &= \left. \mu \frac{\partial}{\partial \mu} \right|_0 u \ , 
\end{align}
we proceed to the flow dependent couplings $\tau(l)$ and $u(l)$
(see Eqs. \ref{tau_HT} and \ref{u_HT})
\begin{align}
l \frac{\partial \tau(l)}{\partial l} &= \tau (l) 
\zeta_\tau(l) \ , \\
l \frac{\partial u(l)}{\partial l} &= \beta_u ( l) \ 
\end{align}
given by the first order ordinary differential equations
\begin{align}
\label{flow_eq1}
l \, \frac{\partial \tau(l)}{\partial l} &=  \tau ( l) 
\left(
-2 +  \frac{n+2}{6} u(l)  \frac{1}{[1+\tau(l)]^{1+ \varepsilon/2}}
\right) \ ,\\
\label{flow_eq2}
l \, \frac{\partial u(l)}{\partial l} &= u ( l) 
\left(
-\varepsilon +  \frac{n+8}{6} u(l)  \frac{1}{[1+\tau(l)]^{1+ \varepsilon/2}}
\right) \ ,
\end{align}
and the initial conditions $\tau(1)=\tau$ and $u(1)=u$.
The asymptotic behavior is determined by zeros of the $\beta$ function, giving
the fixed points of the renormalization group.
Here, we find the Gaussian fixed point $u_G^* = 0$ with $\zeta_G^* = -2$ and the
Heisenberg fixed point $u_H^* =\frac{6 \varepsilon}{n + 8}$ with $\zeta_H^* =
-2+\varepsilon $.
These fixed points 
are of course well-known, \cite{Ami84} but in the generalized minimal
subtraction scheme it is now possible to describe the crossover between these
two fixed points. 
We are interested in the theory in three dimensions and will henceforth 
discuss this case ($\varepsilon =1)$.

\begin{figure}
\epsfig{file=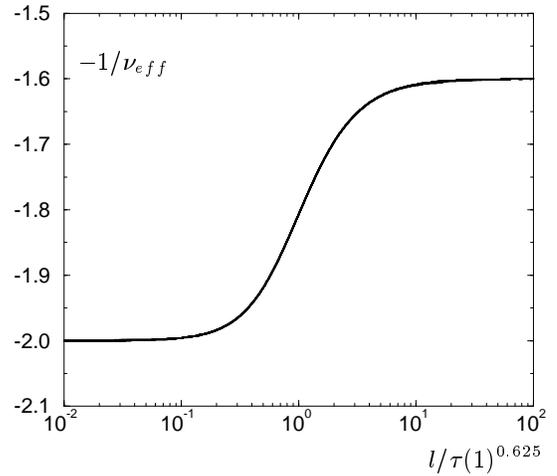, width=8cm} 
\caption{The flow of the renormalized mass parameter 
$\tau(l)$ vs. the scaling variable $x=l/\tau(1)^{0.625}$
for ten different values of $\tau(1)<0.1$ [$u(1)=0.6, n=2$].}
\label{fig1}
\end{figure}

First we investigate the crossover of the $\tau(l)$ flow. 
It is possible to recover
the universal crossover in the flow by plotting $\tau(l)$ against the scaling
variable (compare Refs. \onlinecite{Sch94,Tae92,Tae93}) 
\begin{align}
x = \cfrac{l}{\tau(1)^{1/(2-\frac{n+2}{n+8})}} \ \ .
\end{align}
In Fig. \ref{fig1} the effective exponent for the $x$-dependence of
$\tau(l)$ is depicted for ten different values of $\tau(1)$ [with fixed
$u(1)$ and $n=2$], coinciding perfectly. 
There is a crossover from the region $l \rightarrow 0$
with the exponent $-2$ to the region  $l \rightarrow 1$ with the
exponent $-2 + (n+2)/(n+8)$.

Next we find, with the scaling variable $x \propto (k \xi)^{-1}$,
the effective exponent $\nu_{\text{eff}}$  
of the temperature dependence of the
correlation length 
\begin{align}
\tau(l) \propto l^{-1/\nu_{\text{eff}}} \ \ \Rightarrow \ \ 
\frac{1}{\nu_{\text{eff}}} = 
\begin{cases}
2 & l \rightarrow 0 \\
2 - \frac{n+2}{n+8} & l \approx 1 
\end{cases}  
\end{align}
Thus, with the generalized minimal-subtraction scheme we can describe 
the crossover from the
non-classical critical behavior to the Gaussian behavior, e.g., as a 
function of the temperature variable $\tau$.

\subsubsection{Matching}
To one-loop order, 
only the tadpole graph enters in the two-point function
(see $\Gamma_{\circ \tilde \psi \psi}$ in 
App. \ref{app1}) shifting the
critical temperature as stated above. 
Thus the susceptibility does not change its form and the renormalized version
reads with Eq. (\ref{sus_cor1}), Eq. (\ref{cor_cum1}) and 
App. \ref{app1} to one-loop order,
\begin{align}
\chi_R^{-1} ({\vc k},\omega) = k^2 - i \omega/\lambda + \tau(l) \mu^2 l^2 \ .
\end{align} 
Yet what we gained in the last subsection was the temperature dependence of the
coupling constants. We have to take into consideration this temperature
dependence in order to discuss the changes resulting from the fluctuation 
corrections.
One now has to ask the question: how does the flow enter
the physical quantities measured in an experiment?

With the flow dependence of the coupling constants, the relaxation rate to 
one-loop order becomes [see Eq. (\ref{relax-time})]
\begin{align}
\frac{1}{T_1} &\propto 
               \int_{}^{} k^2 dk \frac{\Im \chi_R({\vc k},\omega_L)}{\omega_L}
               \nonumber \\
              &= \int_{}^{} k^2 dk 
                    \frac{1}{\omega_L^2/\lambda^2 + [k^2 + \tau(l)\mu^2 l^2]^2} \cdot 
                     \frac{1}{\lambda} \nonumber \\
              & =  \frac{1}{\mu l}\frac{1}{\lambda} \int_{}^{} \tilde k^2 d \tilde k 
                 \frac{1}{\tilde \omega_L^2 + [\tilde k^2 + \tau(l)]^2} 
\label{relax1}    
\end{align}
where $\tilde k = k/\mu l$ and $\tilde \omega_L = \omega_L/\lambda \mu^2 l^2$.

\begin{figure}
\epsfig{file=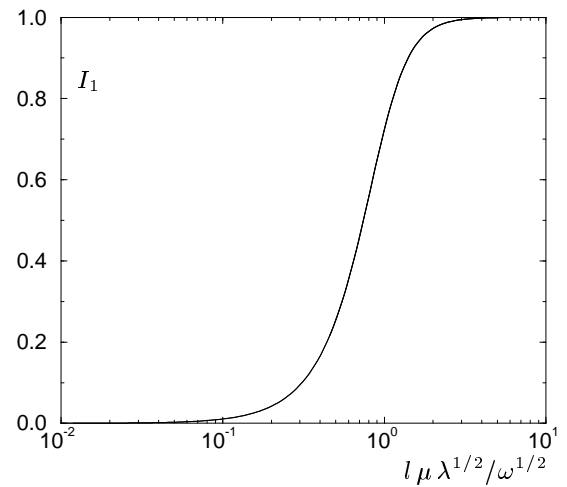, width=8cm} 
\caption{Logarithmic derivative of the integral in Eq. (\ref{relax1}) with 
respect to $l$, plotted vs. the
scaling variable $x=l\mu\lambda^{1/2}/\omega^{1/2}$.}
\label{fig2}
\end{figure}

Keeping $r(l)$ fixed (it is set to 1) the relaxation rate $1/T_1$ is proportional to 
$l^{-1}$ for large $l$. When $l$ approaches zero a constant value of 
the integral and hence of the relaxation rate $1/T_1$ will be reached,
because of the fixed time scale $1/\tilde \omega_L$. 
The physical reason is that in the slow motion limit the 
characteristic time scale ($1/\omega_\varphi$)
becomes larger than the experimental time scale $1/\tilde \omega_L$.
In Fig. \ref{fig2} the logarithmic $l$ dependence of the integral  
\begin{align}
I_1 \equiv \frac{\partial \log T_1}{\partial \log l} = - \cfrac{\partial  \log \left( 
                 \cfrac{1}{\mu l \lambda} \int_{}^{} \tilde k^2 d \tilde k 
                 \cfrac{1}{\tilde \omega_L^2 + (\tilde k^2 + 1)^2} \right) 
                  }{\partial \log l}
\end{align}
is plotted against the scaling variable 
\begin{align}
x = l \cdot \frac{\mu \sqrt{\lambda}}{\sqrt{\omega_L}}\ .
\end{align}
We regain the transition from the $l$-independent regime ($l\rightarrow 0$)
(and therefore of the temperature-independent regime) 
to the regime $I_1 \propto l$ ($l\rightarrow \infty$).
This corresponds to the transition from the slow-motion to the fast-motion
limit; a change of $l$ is aquivalent to a change of $\omega_\varphi$.  

However, we are rather interested in the dependence of the relaxation rate
from the physical temperature $\tau(1)$ than from the flow parameter $l$. 
This may be obtained as follows.
Knowing the solution of the flow equations $\tau(l)$ and $u(l)$,
we can find a $l_1$ for a given $\tau(1)$ that fulfills the equation 
$\tau(l_1)=1$. Inverting this relation, $\tau(1)$ for a given $l_1$ with 
$\tau(l_1)=1$ can be found. 
It is not possible to write down an analytical expression, but numerically this
relationship is readily obtained.
Thus we are led to $1/T_1(r(1)) = 1/T_1(l_1[r(1)])$.
To connect the theory in a region where the perturbation expansion is valid 
with the interesting region, 
we match the temperature variable $r(l)$ to 1,
thus imposing the crossover behavior of the flow $r(l)$ to the effective
exponent of the relaxation time $1/T_1$. 

\begin{figure}
\epsfig{file=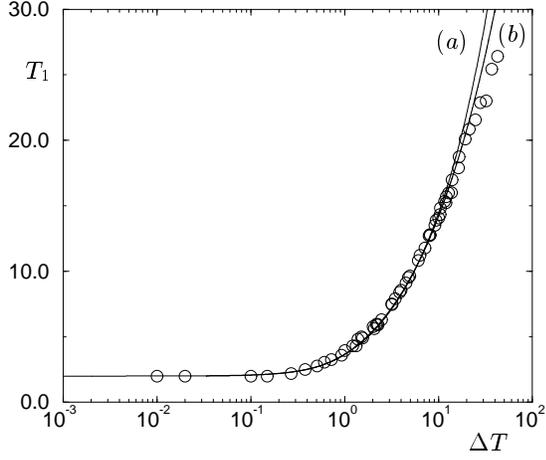, width=8cm} 
\caption{Relaxation time $T_1$ vs. the deviation $\Delta T$ 
from the critical temperature $T_I$. The $NMR$ experimental data,
\cite{Mis97} indicated by circles, are compared with the theoretical results,
represented by the solid lines. A crossover to the mean-field regime starting
at $\Delta T \approx 10 K$ [line (a)] and at $\Delta T \approx 5 K$ [line (b)] is
considered here.}
\label{fig3}
\end{figure}

In Fig. \ref{fig3} the resulting $T_1(T)$ dependence is used to fit 
experimental $NMR$ data for Rb$_2$ZnCl$_4$
(measured points are indicated by circles) from Mischo et al. \cite{Mis97} 
Two parameters have to be fixed in the theory. First, the
prefactor relating the relaxation rate and the integral over the imaginary part
of the susceptibility in Eq. (\ref{relax1}) must be determined. 
Thus, the value of $1/T_1(T=T_I)$ is set. 
The second parameter is the scale of $\omega_L$
compared to the coupling $\lambda \mu$. With this the relative temperature
$\Delta T$, where the transition between slow and fast motion limit takes
place, is adjusted.

The  two fit curves presented in Fig. \ref{fig3} show a crossover 
to the mean-field regime starting at $\Delta T\approx 10K$ (a) 
and at $\Delta T \approx 5K$ (b).  
A very good agreement, not only for the transition from the slow to the fast
motion limit, but also for the high-temperature behavior is found in the
second case.
We want to discuss this issue now in more detail.

\subsubsection{Width of the Critical Region}
Some experiments report large regions 
in which non-classical exponents for the temperature dependence 
of the relaxation rate are observed.
E.g., in Ref. \onlinecite{Hol95} the range above $T_I$ where 
non-classical exponents are found is $\Delta T \approx 100 K$.  
These findings have to be understood by means of the
Ginzburg-Levanyuk \cite{Lev59,Gin60} argument, 
which states that only near to the 
critical temperature the non-classical critical exponents should be valid.
Fluctuations should contribute only near the critical point and change 
the mean-field picture there.

\begin{figure}
\epsfig{file=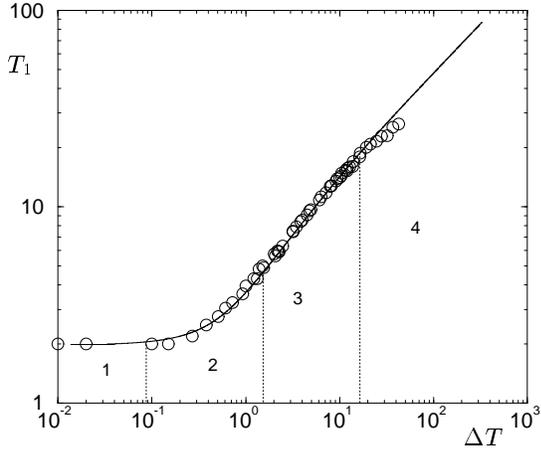, width=8cm} 
\caption{Relaxation time $T_1$ in a logarithmic plot vs. the deviation $\Delta T$ 
from the critical temperature $T_I$. 
The $NMR$ experimental data \cite{Mis97} (circles) are compared 
with the theoretical result, represented by the solid line. 
A crossover to the mean-field regime starting
at $\Delta T \approx 5 K$ is considered.
The four labeled temperature regimes are discussed in the text.}
\label{fig4}
\end{figure}

The property of the integral quantity $1/T_1$, in the region where a
crossover between the non-classical critical exponents and the mean-field
exponents occurs, was studied in the last subsection.
We now comment on the four regions that can be identified and are
indicated by numbers $1 \dots 4$ in Fig. \ref{fig4}.
 
Very close to $T_I$, there is a temperature-independent region (1), because
of the dominating scale $\omega_L$. Here, the probing frequency $\omega_L$ is
too fast to grasp the critical behavior. 
Upon going to higher temperatures, after a transition region (2), 
a temperature dependence with non-classical
critical exponents emerges (3).
For even higher temperatures one finds a crossover to the mean-field
exponents, in regime (4). 
In Fig. \ref{fig4} this crossover takes place between $\Delta T \approx 5K$ 
and $\Delta T \approx 20K$.
From Fig. \ref{fig3}, we find that the crossover at lower temperatures,
here starting at $\Delta T \approx 5K$, 
leads to a better fit of the experimental data.
Thus the reported large region, where supposedly 
non-classical exponents are found,
\cite{Hol95} is in our opinion not an indispensable conclusion that can be
drawn from the experimental data. 
The plausible scenario of an extended crossover regime beyond the truly
asymptotic region of width 
$\Delta T \approx 5K$ is in fact in perfect agreement with the data.

As this is not a universal feature other scenarios are possible. It may happen 
that the scale of $\omega_L$ is very large and thus only the Gaussian exponents 
are found.  
We omitted the contribution of higher Raman processes, as discussed by Holzer
et al., \cite{Hol95} leading to an additional $T^2$ dependence 
for the relaxation rate. These would bend the curves downward even more and
explain the deviation present at the highest measured temperatures. 
Not taking these additional contributions into consideration,
however, clarifies the crossover aspect.  

\section{Low-temperature phase}
This section is devoted to the incommensurate ordered phase 
below the critical temperature $T_I$.
In the $O(n)$-symmetrical model a spontaneous breaking of a
global continuous symmetry occurs and the expectation value of the order 
parameter becomes nonzero. 
Now parallel and perpendicular fluctuations with respect to the nonzero order 
parameter have to be distinguished. 
As a consequence there appear $n-1$ massless Goldstone modes
which lead to infrared singularities for all temperatures below $T_I$ in certain
correlation functions. \cite{Nel76,Maz76} 
We investigate how these Goldstone modes influence
the dynamical properties of the quantities we are interested in, e.g., 
the $NMR$ relaxation rate.
To do so, we first 
derive the dynamical functional appropriate below $T_I$. In the following
subsection some comments about the Goldstone anomalies are made. We will then
treat the dynamics of the fluctuations parallel (amplitudons) and perpendicular
(phasons) to the order parameter.
Again a renormalization group calculation to one-loop order is presented. We
will discuss the dynamical susceptibility before evaluating the integrals
leading to the relaxation rate. In the last section, we compare 
with experimental data. 
We also comment on the existence of a phason gap.

\subsection{Dynamical functional}
Let us assume that the spontaneous symmetry breaking below $T_I$ 
appears in the $n$th direction of the order parameter space. As usual, new
fields $\pi_\circ^\alpha$, $\alpha = 1,\ldots,n-1$, and 
$\sigma_\circ$ are introduced \cite{Law81} 
\begin{align} 
\binom{{\tilde \psi}_\circ^\alpha}{{\tilde \psi}_\circ^n} = \binom{{\tilde
\pi}_\circ^\alpha }{ {\tilde \sigma}_\circ} \quad , \qquad
\binom{\psi_\circ^\alpha }{ \psi_\circ^n} = \binom{\pi_\circ^\alpha }{ \sigma_\circ
+ {\bar \phi}_\circ} \quad , 
\end{align}
with 
\begin{align}
\langle \pi_\circ^\alpha \rangle  = \langle \sigma_\circ \rangle = 0 \ .
\label{expec_fluc}
\end{align}
The order parameter is parameterized as
\begin{align}
{\bar \phi}_\circ = \sqrt{\frac{3 }{ u_\circ}} \, m_\circ \ .
\label{OP_tktc}
\end{align}
Thus $\sigma_\circ$ corresponds to the longitudinal,
and $\pi_\circ^\alpha$ to the transverse fluctuations.

\begin{figure}
\epsfig{file=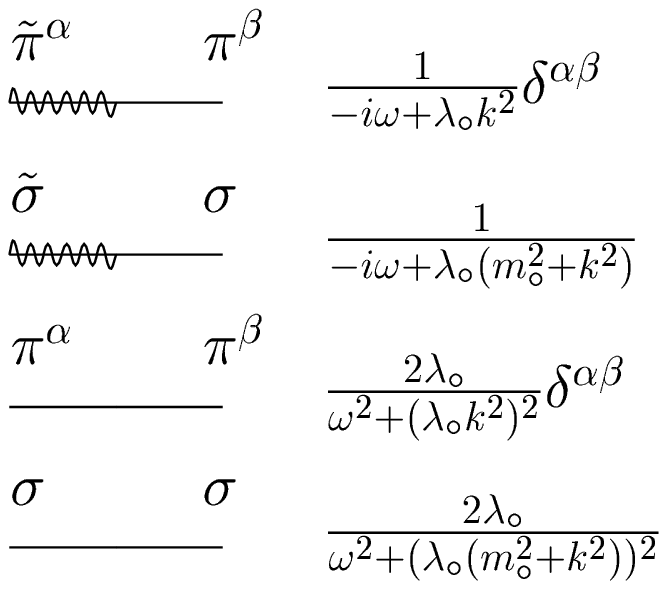, width=6.0cm} 
\caption{Propagators below $T_I$.}
\label{fig6}
\end{figure}

\begin{figure}
\epsfig{file=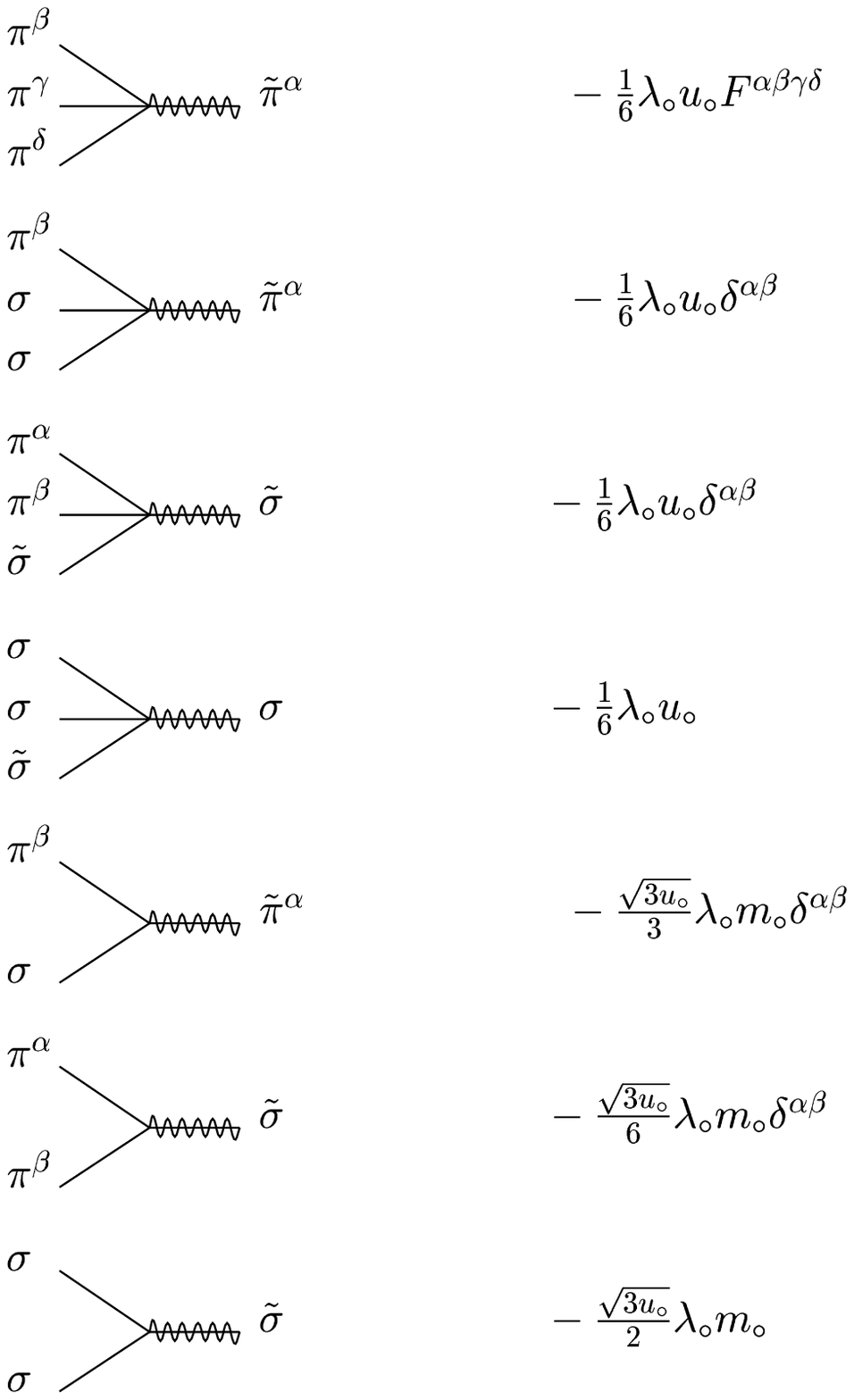, width=9.5cm} 
\caption{Vertices  below $T_I$.}
\label{fig7}
\end{figure}

Inserting these transformations into the functional (\ref{dyn_func})
leads to a new functional of the form $ J = J_0 + J_{int} + J_1+ const$
with \cite{Tae92}
\begin{align}
J_0 & [\{  {\tilde \pi}_\circ^\alpha \} , {\tilde \sigma}_\circ 
, \{ \pi_\circ^\alpha \} , \sigma_\circ] = \nonumber \\
&  \int_k \int_\omega \biggl[
\sum_\alpha  \lambda_\circ \,  \, {\tilde \pi}_\circ^\alpha({\vc
k},\omega) \, {\tilde \pi}_\circ^\alpha(- {\vc k},-\omega) \nonumber 
\\
& \qquad \qquad +  \lambda_\circ  \, {\tilde \sigma}_\circ({\vc k},\omega) \, {\tilde
\sigma}_\circ(- {\vc k},-\omega) \nonumber \\
& - \sum_\alpha {\tilde \pi}_\circ^\alpha({\vc k},  \omega) \, \Bigl[ i
\omega + \lambda_\circ \,  \, \Bigl( r_\circ + \frac{m_\circ^2}{ 2} + k^2
\Bigr) \Bigr] \, \pi_\circ^\alpha(- {\vc k},- \omega) \nonumber \\
&- {\tilde \sigma}_\circ({\vc k}, \omega) \, \Bigl[ i \omega +
\lambda_\circ \,  \, \Bigl( r_\circ + \frac{3 \, m_\circ^2 }{ 2} + k^2
\Bigr) \Bigr] \, \sigma_\circ(- {\vc k},- \omega) \biggr] \ ,
\end{align}
\begin{align}
&  J_{int} [\{ {\tilde \pi}_\circ^\alpha \} , {\tilde
\sigma}_\circ , \{ \pi_\circ^\alpha \} , \sigma_\circ] = \nonumber \\ 
& - \frac{1 }{ 6} \,
\lambda_\circ \, u_\circ \int_{k_1 k_2 k_3 k_4} \int_{\omega_1 \omega_2
\omega_3 \omega_4} \delta \! \left( \sum_i {\vc k}_i
\right) \, \delta \! \left( \sum_i \omega_i \right) \nonumber \\
& \times \biggl[ \sum_{\alpha \beta} {\tilde
\pi}_\circ^\alpha({\vc k}_1,\omega_1) \, \pi_\circ^\alpha({\vc
k}_2,\omega_2) \, \pi_\circ^\beta({\vc k}_3,\omega_3) \,
\pi_\circ^\beta({\vc k}_4,\omega_4) \nonumber \\
& \quad + \sum_\alpha {\tilde \pi}_\circ^\alpha({\vc
k}_1,\omega_1) \, \pi_\circ^\alpha({\vc k}_2,\omega_2) \,
\sigma_\circ({\vc k}_3,\omega_3) \, \sigma_\circ({\vc k}_4,\omega_4)
\nonumber \\ 
& \quad + \sum_\alpha {\tilde \sigma}_\circ({\vc
k}_1,\omega_1) \, \pi_\circ^\alpha({\vc k}_2,\omega_2) \,
\pi_\circ^\alpha({\vc k}_3,\omega_3) \, \sigma_\circ({\vc
k}_4,\omega_4) \nonumber \\
&\quad +  {\tilde \sigma}_\circ({\vc
k}_1,\omega_1) \, \sigma_\circ({\vc k}_2,\omega_2) \, \sigma_\circ({\vc
k}_3,\omega_3) \, \sigma_\circ({\vc k}_4,\omega_4) \biggr]  \nonumber \\  
& - \lambda_\circ \, {\sqrt{\frac{3 \, u_\circ}{ 6}}} \, m_\circ \int_{k_1 k_2
k_3} \int_{\omega_1 \omega_2 \omega_3} \delta \! \left(
\sum_i {\vc k}_i \right) \, \delta \! \left( \sum_i \omega_i
\right) \nonumber \\
& \times \biggl[ \sum_\alpha 2 \, {\tilde \pi}_\circ^\alpha({\vc
k}_1,\omega_1) \, \pi_\circ^\alpha({\vc k}_2,\omega_2) \,
\sigma_\circ({\vc k}_3,\omega_3) \nonumber \\
& \quad + \sum_\alpha {\tilde \sigma}_\circ({\vc k}_1,\omega_1) \,
\pi_\circ^\alpha({\vc k}_2,\omega_2) \, \pi_\circ^\alpha({\vc k}_3,\omega_3) 
\nonumber \\ 
& \quad + 3 \, {\tilde \sigma}_\circ({\vc k}_1,\omega_1) \, 
\sigma_\circ({\vc k}_2,\omega_2) \, \sigma_\circ({\vc k}_3,\omega_3)
\biggr] \ ,
\end{align}
and
\begin{align}
J_1[{\tilde \sigma}_\circ] = & 
- \lambda_\circ \,  \sqrt{\frac{3}{ u_\circ}} \,
m_\circ  \, \left( r_\circ + \frac{m_\circ^2 }{ 2} \right) \, \nonumber \\
& \qquad \times \, \int_k \int_\omega
 \, {\tilde \sigma}_\circ(-{\vc k},-\omega) \delta ( {\vc k}, \omega) \ .
\end{align}
Equation (\ref{expec_fluc}) 
($\langle \sigma_\circ \rangle = 0$) yields a perturbative identity
that gives the relation between $r_\circ$ and $m_\circ$,
reading to one-loop order \cite{Law81}
\begin{align}
r_\circ + \frac{m_\circ^2 }{ 2} = &  
- \frac{n - 1}{ 6} \, u_\circ \int_k \frac{1 }{ r_\circ + 
\frac{m_{\footnotesize 0}^2 }{
2} + k^2}  \nonumber \\
& - \frac{1 }{ 2} \, u_\circ \int_k \frac{1 }{ r_\circ + \frac{3 \, 
m_{\footnotesize 0}^2 }{ 2} + k^2} \ .
\end{align}
In the following $r_\circ$ is replaced by $m_\circ$. Notice that
by using the variable $m_\circ$, the shift of $T_I$ is already incorporated 
[see Eq. (\ref{OP_tktc})].
We can now write down the basic ingredients needed to apply the recipe for the
dynamical perturbation theory below $T_I$. The emerging propagators, vertices,
and counterterms are listed with their graphical
representation in Figs. \ref{fig6}, \ref{fig7}  and \ref{fig8} (see Ref.
\onlinecite{Tae92}).

\subsection{Goldstone theorem and coexistence limit}
As mentioned before, the particularity of the $O(n)$-symmetric
functionals below the critical temperature is the occurrence of Goldstone 
modes in the entire low temperature phase. \cite{Gol61,Wag66,Nel76}  
Because no free energy is required for
an infinitesimal quasistatic rotation of the order parameter, the transverse
correlation length diverges in the limit of zero external field. 
The corresponding massless modes are the Goldstone modes, \cite{Gol61}
in this context called phasons.
They are manifest in non-analytical behavior of correlation functions, 
for example the longitudinal static 
susceptibility diverges and  
changes its leading behavior from being proportional to $k^{-2}$ to
\cite{Nel76,Maz76,Law81,Tae92} 
\begin{align}
\chi_L^{-1}({\vc k},0) \propto k^\varepsilon \ .
\end{align}
Before discussing the details of the renormalization theory below $T_I$, 
we summarize some important aspects, which explain
why below $T_I$ an $\varepsilon$-expansion can be avoided.
For more details see Ref. \onlinecite{Tae92}. 

Leaving the critical temperature region $T \approx T_I$, 
which is characterized in the non-perturbed case 
by $m_\circ =0$, and lowering the temperature, means
that the fluctuations of the longitudinal modes (amplitudons) 
become negligible, because these modes are massive ($m_\circ$). 
In contrast 
the phasons remain massless and hence their fluctuations will dominate.

Yet a different way of describing the 
dominance of the fluctuations of the Goldstone modes is to consider
the spherical model limit \cite{Maz76} $n \rightarrow \infty$.
In this case of ``maximal" symmetry breaking, the Goldstone modes are weighted
with the factor $n-1 \rightarrow \infty$.

As mentioned above, 
below $T_I$ coexistence anomalies are present. They arise from the fact that
the $n-1$ transverse modes are massless. In the limit $k \rightarrow 0$
and $\omega \rightarrow 0$ for $T<T_I$ this manifests itself 
in typical infrared
divergences. An important point to remember is that in order
to gain these coexistence anomalies, one can also study the case $m_\circ
\rightarrow \infty$.
In the renormalization scheme it is shown that the flow of
the mass parameter $m_\circ$ tends to infinity as the 
momentum and frequency tend to zero. 

From this it is plausible, and was also proved \cite{Law81},
that in the coexistence limit the result for the
two-point vertex functions are identical with the results arising from the 
spherical model limit $n \rightarrow \infty$.

These findings render an $\varepsilon$ expansion unnecessary in the
coexistence limit ($k\rightarrow0$, $\omega \rightarrow 0$ at $m_\circ>0$), 
because the
asymptotic theory (the spherical model) is exactly treatable and reduces to the
zero- and one-loop contributions. 
Of course, one has to make sure that the
properties of the asymptotic functional will be reproduced in the respective
limit. 
Within the generalized minimal subtraction scheme this is possible. As stated
in subsection \ref{tgtc_renorm} Lawrie's method \cite{Law81}
and its dynamical extension 
in Ref. \onlinecite{Tae92} lead beyond these limits 
and allow for a detailed study of the crossover behavior.  

The behavior of the correlation functions in our case is 
driven by the crossover between
the three fixed points present below $T_I$. 
Besides the Gaussian fixed point one finds the Heisenberg fixed point 
\begin{align}
[u_H = 6 \epsilon/(n+8)] 
\end{align}
and the coexistence fixed point \cite{Law81} 
\begin{align}
[u_C=6\epsilon/(n-1)] \ .
\end{align}
We will again employ the generalized minimal subtraction scheme to study the 
crossover between these fixed points.

\subsection{Renormalization group analysis below \boldmath{$T_I$}}
\subsubsection{Flow equations}
Below $T_I$, using only the one-loop diagrams, again the field
renormalization vanishes. 
Hence, the only non-trivial $Z$ factors are 
the ones for the temperature scale and the coupling constant:
\begin{align}
m^2 &= Z_m^{-1} m_\circ^2 \mu^{-2} \ , \\
u &= Z_u^{-1} u_\circ A_d \mu^{-\varepsilon} \ .
\end{align}
Because we use $m$ instead of $r$ an important relationship can be stated,
which is true independently of the loop order, \cite{Tae92}
\begin{align}
Z_m \cdot Z_\sigma = Z_u \ .
\label{Zum}
\end{align}
To one-loop order ($Z_\sigma = 1$) we find (see App. \ref{app1}) \cite{Tae92}
\begin{align}
Z_u = Z_m =&  1 + \frac{n-1}{6 \varepsilon} u_\circ A_d \mu^{- \varepsilon}
\nonumber \\
& + \frac{3}{2 \varepsilon} u_\circ A_d \mu^{- \varepsilon} 
\frac{1}{(1+m_\circ^2/\mu^2)^{\varepsilon/2}} \ .
\label{Z_kt}
\end{align}
Here, the contribution of the transverse loops lead to different divergences
manifest in the change of the $Z$ factors compared to those above $T_I$ [see
Eq. (\ref{Z_gt})].
We recover the familiar renormalization constant in the critical region by
setting $m_\circ=0$. When considering the coexistence limit $m_\circ
\rightarrow \infty$, the weight of the effective critical fluctuations is
reduced from $n+8$ to $n-1$, the number of Goldstone modes. 

Asymptotically ($m \rightarrow \infty$)
the $Z$ factors are exact. In the crossover region they are an approximation to
the order $u_\circ^2/(1+m_\circ/\mu^2)^{\varepsilon/2}$. \cite{Tae92}

From this we directly derive the flow-dependent couplings 
\begin{align}
l \, \frac{\partial m(l)}{\partial l} = & \frac{1}{2} m ( l) 
\left( -2 +  \frac{n-1}{6} u(l)   \right.  \nonumber \\ 
& \left.  + \frac{3}{2} \frac{u(l)}{[1+m(l)^2]^{1+ \varepsilon/2}} \right)  \ ,
\\
l \, \frac{\partial u(l)}{\partial l} = & u ( l) 
\left(
-\varepsilon +  \frac{n-1}{6} u(l)  \right. \nonumber \\
& \left.  + \frac{3}{2}\frac{u(l)}{[1+m(l)^2]^{1+ \varepsilon/2}} \right) \ .
\end{align}
Wilson's flow equations $\beta_u$ and $\zeta_m$ now read
\begin{align}
l \frac{\partial m(l)}{\partial l} &= \frac{1}{2} m(l) \zeta_m(l) 
\ , \\ 
l \frac{\partial u(l)}{\partial l} &= \beta_u(l) \ .
\end{align}
Three fixed points have now to be taken into consideration. \cite{Law81,Tae92}
Next to the Gaussian fixed point $u_G^* =0$ with $\zeta_{mG}^* = -2$, we find
in the critical limit ($m_\circ \rightarrow 0$)
the infrared-stable Heisenberg fixed point $u_H^* =6\varepsilon/(n+8)$
with $\zeta_{mH}^* = -2+\varepsilon$. In the coexistence limit $m_\circ 
\rightarrow \infty$, we find in addition to the still 
ultraviolet-stable Gaussian fixed
point the coexistence fixed point, identified by Lawrie,
\cite{Law81} $u_C^* =6\varepsilon/(n-1)$ with $\zeta_{mH}^* =
-2+\varepsilon$, which is infrared-stable.
Thus $m(l)^2$ diverges asymptotically for $l\rightarrow 0$ as
$l^{-2+\varepsilon}$, if $\varepsilon < 2$. Indeed, the coexistence limit is 
described by a divergent mass parameter.

In Figs. \ref{fig10} and \ref{fig11} the flow for $m(l)$ and $u(l)$ is
plotted. We find for the flow $u(l)$ 
a crossover between the coexistence fixed point, inversely 
proportional to the number of Goldstone modes $(n-1)$, and the Heisenberg
fixed point  
\begin{align}
u(l) \ \ \Rightarrow \  
\begin{cases}
6 \varepsilon/(n-1) & l \rightarrow 0 \\
6\varepsilon /(n+8) & l \approx 1 \ .
\end{cases} 
\end{align}
That means that for $m(1) \ll 1 $ the coexistence limit is not 
approached directly for
$l\rightarrow 0$, but for a while the flow stays near 
the Heisenberg fixed point regime. 
The scaling variable for $u(l)$ is here \cite{Sch94,Tae92}
\begin{align}
x=\frac{l}{m(1)^{2/(2-\varepsilon)}} \ ,
\end{align}
again leading to perfectly coinciding curves when plotted vs. $x$.

From the relation stated in Eq. (\ref{Zum}) 
one can deduce the renormalization-group invariant \cite{Tae92}
\begin{align}
\frac{m(l)^2}{u(l)} l^{2-\varepsilon} = \frac{m(1)^2}{u(1)} \ ,
\end{align}
which immediately gives us the scaling of $m(l)$ that can be observed in
Fig. \ref{fig10}, 
\begin{align}
m(l)^2 \propto l^{-1/\nu_{\text{eff}}} \ \ \text{with} \ \ 
\frac{1}{\nu_{\text{eff}}} = 
\begin{cases}
2-\varepsilon & l \rightarrow 0 \\
2- \frac{n+2}{n+8} & l \approx 1 \ .
\end{cases} 
\end{align}
Notice that the value of $1/\nu_{\text{eff}}$
in the first case  is the same as $1/\nu$ for the spherical model.
The mass parameter $m$ diverges for $l \rightarrow 0$, $m(l)^2 \propto
l^{-2+\varepsilon}$ with $\varepsilon<2$. 

From now on we will concentrate on three dimensions ($\varepsilon =1$) and
$n=2$. 

\subsubsection{Matching}
In order to discuss the static susceptibility we use the matching 
condition $\mu l=k$. 
This relation connects the dependence of
the renormalized quantities on the momentum scale $\mu$ 
with the $k$ dependence, in which we are interested.

For the integrated value, we are interested in the 
temperature dependence rather than the dependence 
on the flow parameter $l$ or the wavevector $k$. 
Thus, after integration we have to again
match the resulting $l$-dependent
relaxation rate with the physical temperature (see section
\ref{tgtc_renorm}).

\subsection{Susceptibility}
In order to determine the renormalized dynamical susceptibility, we evaluate
the one-loop diagrams for the two-point cumulants, which can be easily derived
from the one-loop vertex functions listed in Appendix \ref{app1}. \cite{Tae92}
Below $T_I$, the structure of the susceptibility does change to one loop order,
compared to the mean-field results. 
We write in a general form
\begin{align}
\chi_{\circ \ \perp / \parallel}^{-1}( {\vc k}, \omega)  
= \frac{- i \omega}{\lambda_\circ} + k^2 + 
f_\circ^{\perp / \parallel} ( {\vc k}, \omega)  \ ,
\end{align}
with the self-energy $f_\circ^{\perp / \parallel}$ containing the
contributions of the one-loop diagrams. 
The explicit form of $f_\circ^{\perp / \parallel}$ is gained from the 
calculation of the two-point vertex function in App. \ref{app1} and Eqs.
(\ref{sus_cor1}) and (\ref{cor_cum1}).
We then obtain the renormalized susceptibility $\chi_{\parallel,\perp}^R$
by inserting the $Z$
factors with the flow dependent coupling constants $u(l)$ and $m(l)$.
Because no field renormalization is present, we can replace $\lambda_\circ$
with $\lambda$ to this order. 
This is because via the fluctuation-dissipation theorem 
(\ref{fluk_diss}), $Z_\lambda$
and the field renormalizations are connected.

\endmcols

\subsubsection{Amplitudon modes}

In $d=3$ the longitudinal susceptibility characterizing 
the amplitudon modes is given by \cite{Tae92}
\begin{align}
\chi_{\parallel}^R ( {\vc k}, \omega)  =&
\, \frac{1}{ k^2 - i \,
\omega / \lambda + \mu^2 \, l^2 \, m(l)^2 \, Z_m(l)}
\Biggl( 
 1 \nonumber 
 + \frac{1 }{ (k^2 - i \, \omega / \lambda) / \mu^2 \,l^2 
+ m(l)^2} \, \frac{u(l) \, m(l)^2 }{ 2 \, k / \mu
\, l} \nonumber \\
& \times \Biggl[ \frac{n - 1 }{ 3} \left( \frac{\pi
}{ 2} + \arcsin \frac{i \, \omega / \lambda }{ k^2 - i \, \omega
/ \lambda} \right) \nonumber 
+ 3 \Biggl( \arcsin \frac{ i \, \omega / \lambda \,
\mu^2 \, l^2 }{ \left[ \left( [k^2 - i \, \omega / \lambda
] / \mu^2 \, l^2 \right)^2 + 4 \, m(l)^2 \, k^2/
\mu^2 \, l^2 \right]^{1/2} } \nonumber \\
& + \arcsin \frac{ (k^2 - i \, \omega / \lambda)/
\mu^2 \, l^2 }{ \left[ \left( [k^2 - i \, \omega / \lambda
] / \mu^2 \, l^2 \right)^2 + 4 \, m(l)^2 \, k^2 / \mu^2 \, l^2 
\right]^{1/2} } \Biggr) \Biggr] \Biggr)
\label{chi_long}
\end{align}
\beginmcols
First we want to discuss some limits in order to become acquainted with
this complex form of the susceptibility. 

It is important to notice the change of the structure of the $RG$ susceptibility 
that results from the one-loop contribution of the perturbation theory.
To clarify this, we state the asymptotic susceptibility, which is
evaluated for non-zero frequency
$({\vc k} \rightarrow 0, \omega > 0)$
\begin{align}
\chi_{\parallel}^{R} = \cfrac{1}{k^2 -i \omega/\lambda + 
m(1)^2 \cdot k \cdot \cfrac{1}{1+ a \cdot g(k)}} 
\end{align}
with a constant $a$ and a function $g(k)$ that is regular for $q
\rightarrow 0$. 
From this limit, it becomes clear that we have to expect changes of
the scaling behavior.

In the coexistence limit $(\omega=0, {\vc k} \rightarrow 0)$ 
we recover the exact asymptotic result ($d=3$, $\varepsilon=1$) 
\begin{align}
\chi_{\parallel}^{R} \propto k^{-1}
\label{coex_trans}  
\end{align} 
displaying the coexistence anomaly.
When keeping the frequency $\omega > 0$ fixed 
$({\vc k} \rightarrow 0, \omega \neq 0)$ the imaginary part of
the susceptibility approaches a constant value
\begin{align}
\chi_{\parallel}^{R} \Rightarrow h(\omega) \ .  
\end{align} 
where $h(\omega)$ is a function of $\omega$ only. 
We can now turn to the full susceptibility.
The imaginary part of the susceptibility is plotted for
different temperatures in Fig. \ref{fig13}. 
The structure of $\Im \chi_{\parallel}^{R}$ 
changes dramatically as compared to the mean-field result, as to be expected.
The contributions of the phason and amplitudon loops are given by the terms in
brackets of Eq. (\ref{chi_long}). 
They give rise to a qualitatively different behavior of 
$\chi_{\parallel}^{R}$.
Different scaling regions can be identified. Expanding the imaginary part of 
$\chi_{\parallel}^{R}$ yields analytical expressions for the scaling
regions, as listed in Table \ref{tab2}.
While the $k \rightarrow \infty$ and $k \rightarrow 0$ behavior reproduces
the mean-field result, the correct treatment of the Goldstone anomalies lead to
an additional $k^{-3}$ behavior in the intermediate region
$\sqrt{\omega/\lambda} < k < m(1)$.
A plateau appears for smaller $k$ and temperatures far away from the
critical temperature $T_I$.
The effective exponent $\kappa$ of the $k$ dependence of $\Im
\chi_{\parallel}^{R}$ is plotted in Fig. \ref{fig14}.
One can therefrom easily identify the scaling regions presented 
in Table \ref{tab2}. 

The influence of the Goldstone modes is therefore to alter the $k$ dependence
of the susceptibility not only in the coexistence limit, but also in
intermediate regions. In order to derive 
the temperature dependence of $\chi_\parallel^R$,  in
addition the flows of $m(l)$ and $u(l)$ need to be considered.

\endmcols

\subsubsection{Phason modes}

For the transverse susceptibility characterizing the phason modes one finds
\cite{Tae92}
\begin{align}
\chi_{\perp}^R ( {\vc k}, \omega)   = &
 \frac{1}{ k^2 - i \, \omega / \lambda} 
\Biggl( 1   - \frac{u(l) \, m(l) / 6}{(k^2 - i \, \omega / \lambda) / 
\mu^2 \, l^2} 
\Biggl[ 2 - \frac{m(l)}{k / \mu \, l} 
\Biggl(  \frac{\pi}{ 2} 
- \arcsin \frac{- i \, \omega / \lambda \, \mu^2 l^2 + m(l)^2}{
(k^2 - i \, \omega / \lambda ) / \mu^2 \, l^2 + m(l)^2} 
\nonumber \\
&+  \arcsin \frac{i \, \omega / \lambda \, \mu^2 \, l^2 + m(l)^2 }{ 
\left[ \left( 
(k^2 - i \, \omega / \lambda) / \mu^2 \, l^2 - m(l)^2 \right)^2 + 
4 \, m(l)^2 \, k^2 / \mu^2 \, l^2  \right]^{1/2} }  \nonumber \\
& +  \arcsin \frac{(k^2 - i \, \omega) / \lambda \mu^2 \, l^2 
- m(l)^2 }{ \left[ \left( 
(k^2 - i \, \omega / \lambda ) / \mu^2 \, l^2 - m(l)^2 \right)^2 + 
4 \, m(l)^2 \, k^2 / \mu^2 \, l^2 \right]^{1/2} }  \Biggr) \Biggr]
\Biggr) \ .
\label{chi_tran}
\end{align}
\beginmcols
Here the problem lies in the cancellation of terms with
respect to their $k$ dependence, hidden in the complex structure of
Eq. (\ref{chi_tran}).
Hence we start again with considering the coexistence limit
($ {\vc k} \rightarrow 0, \omega \rightarrow 0$):
\begin{align}
\chi_{\perp}^{R} \propto k^{-2} \ ,
\end{align}
which is easily found.
The results for  $ {\vc k} \rightarrow 0, \omega \neq 0$ are more difficult to
obtain, because the $\arcsin$-terms cancel 
their $k$-dependence against each other. In two limits this can be
done analytically.  
For $m \rightarrow 0 $ one gets 
\begin{align} 
\chi_{\perp}^{R} \rightarrow 
\chi_{\circ \perp} = \frac{1}{k^2 -i\omega / \lambda}
\end{align}
reproducing the mean-field susceptibility for the massless transverse modes.

For $m \rightarrow \infty$ the $\arcsin$-terms read as
$\frac{2}{m} + c_1 \frac{1}{m^3} + c_2 \frac{ i \omega}{\lambda k^2}
\cdot  \frac{1}{m^3}$ leading to 
\begin{align}
\chi_{\perp}^{R} \rightarrow 
 \cfrac{1}{k^2 -i\omega / \lambda + \cfrac{1}{6} u(l)(c_1 k^2 + c_2
\frac{i\omega}{\lambda})/m(l)}
\end{align}
Here $c_1$ and $c_2$ are constants.
Thus in the two extreme limits, the temperature dependence vanishes, 
and only in between,
for $m(l) \propto {\cal O} (1)$ can we expect a slightly 
temperature-dependent behavior.

In Figs. \ref{fig15} and \ref{fig16} the imaginary part of the
transverse susceptibility and its effective exponent with respect to $k$ are
plotted for different temperatures. 
Notice that leaving the
critical temperature leads to a temperature dependence. 
This is caused by the coupling of the amplitudon and phason modes.
Yet as the temperature is further reduced, 
the temperature dependence disappears again.

\subsection{Relaxation rate}

In this subsection we study consequences for the relaxation arising
from the Goldstone anomalies present in the susceptibility. 
As mentioned in Sec. \ref{sec_nmr}, in order to gain the relaxation rate 
we have to integrate over the imaginary part of the susceptibility.

Because the transverse susceptibility is temperature dependent, also the
relaxation rate, connected with the phasons, will be temperature dependent.
This is of course not the case in the mean-field analysis.
As discussed in the last section, for $T \rightarrow T_I$  
the susceptibility approaches the mean-field
result, and thus the relaxation rate at the critical temperature is 
unaltered.

For the relaxation rate connected with the amplitudons the changes are more
subtle. 
Therefore, we collect all the contributions from the one-loop diagrams in a
function $f({\vc k},\omega)$, which  can be interpreted as a 
${\vc k}$- and $\omega$-dependent dimensionless self-energy. 
The susceptibility has now the following
structure 
\begin{align}
(\chi_\parallel^R)^{-1} = k^2 - i\omega/\lambda + f({\vc k}, \omega) \mu^2 l^2 
\end{align}
The dependence of $f$ on $k$ and
$\omega$ is plotted in Fig. \ref{fig17}.
We see that the real part of the effective mass $f$ is decreasing for 
$k \rightarrow 0$. Thus, the 
Goldstone anomalies lead to a reduction of the real part of 
$f$. In the coexistence limit, $\omega \rightarrow 0$ and small $k$,  
the real part of $f$ tends to 0 linearly and 
relation (\ref{coex_trans}) is recovered.
The imaginary part is only $k$-dependent for very small $k$. 
That means, when we  integrate over all $k$ the influence of the Goldstone
anomalies can be interpreted as follows.
The effective Larmor frequency is raised and the mass
is lowered for small $k$ as compared to the mean-field description. 
We can easily derive this easily from the longitudinal
relaxation rate with $f(k,\omega)$ taken into consideration
\begin{align}
  \frac{1}{T_1^\parallel}  \propto & 
               \int_{}^{} k^2 dk \frac{\Im \chi_\parallel^R ({\vc k},\omega_L)}{\omega_L}
               \nonumber \\
              = & \int_{}^{} k^2 dk 
                    \frac{1}{\mu^2 l^2 \omega_L} \nonumber \\
                  &  \times \, \frac{\tilde \omega - \Im
                    f^\parallel(\tilde {\vc k},\tilde \omega)} 
               {[\tilde \omega^2 - 
                     \Im f^\parallel(\tilde {\vc k},\tilde \omega)]^2 + [\tilde
                     k^2 + 
                     \Re f^\parallel(\tilde {\vc k},\tilde \omega)]^2} 
                     \nonumber \\
              = &  \frac{1}{\mu l}\frac{1}{\lambda} \int_{}^{} \tilde k^2 d \tilde k 
                 \nonumber \\
             & \times \, \frac{1- \Im f^\parallel(\tilde {\vc k},\tilde \omega)/{\tilde
                 \omega}}{[\tilde \omega - \Im f^\parallel(\tilde {\vc
                 k},\tilde \omega) ]^2 + [\tilde k^2 +  \Re f^\parallel(\tilde
                 {\vc k},\tilde \omega)]^2}  \ ,    
\end{align}
where again $\tilde k = k/\mu l$ and $\tilde \omega = \omega_L/\lambda \mu^2
l^2$. When we compare this result to the mean-field result, the interpretation
given above becomes clear. The relaxation rate is raised through the
influence of the Goldstone modes. Both the transverse and longitudinal
relaxation times are plotted in Fig. \ref{fig18}. 
Again we have compared our findings with experimental data, 
taken from Ref. \onlinecite{Mis97}. 
In the low-temperature phase we have less freedom of choice in our theory,
as the scale $T_1(T=T_I)$ and the parameters are already 
fixed by their high-temperature values. Thus only one parameter is left to
be adjusted. 

In the vicinity of $T_I$ we find a temperature-independent region, 
because both the
transverse and the longitudinal susceptibility become temperature-independent
and approach their mean-field values.

The transverse relaxation time shows a slight temperature dependence for
temperatures further away from $T_I$. If we use the identical choice of
parameters as for the high-temperature phase, 
we find good agreement in the low-temperature phase as well. 
The temperature where the maximum value of the transverse 
relaxation time in our theory is reached is identified with the corresponding 
temperature in the experiment.
This temperature dependence is due to the coupling between the phason and
amplitudon modes. We want to emphasize that, in agreement with the analysis of
ultrasonic attenuation experiments, \cite{Sch92,Sch94} no phason gap has to be
introduced to explain the experimental data for Rb$_2$ZnCl$_4$.
However, it is important to treat the influence of the Goldstone modes 
beyond the mean-field approach.

For the longitudinal relaxation time, the crossover temperature represents
additional an important scale.
We used the same range $\Delta T \approx 5K$ as in the 
high-temperature phase for the plot in Fig. \ref{fig18}. 
Again good agreement between experiment and theory is observed. 
Both theory and experiment show two scaling regions, one above 
$\Delta T \approx 5K$ and one below. 
The qualitative behavior is correctly reproduced, 
but the quantitative agreement for the 
longitudinal relaxation rate is not as good as compared to the high-temperature 
phase. 
A possible reason may be the following.
We calculated the coupling of the transverse and longitudinal
order parameter fields to one-loop order.
Below $T_I$, the coupling of the order parameter is changing the susceptibility 
in its structure, whereas above $T_I$ nothing dramatic happens.
One has to expect that below $T_I$ this is of course only the first step
beyond mean-field theory and the two-loop corrections might lead to
quantitative modifications in the crossover region.   
Comparing the calculated 
transverse and the longitudinal relaxation rates below $T_I$
with the experimental data is in agreement with this.
A slight temperature dependence is not as sensitive as the scaling behavior of
the longitudinal relaxation time, which has again two regimes due to the
crossover scenario. 

As the  characteristic features are reproduced correctly, we may say that
we understand the complex temperature dependence below $T_I$ in the context of
the coupled order parameter modes and a careful treatment of the Goldstone
modes. 
Upon introducing
the $k$- and $\omega$-dependent self energy $f$ we could in addition provide
a physical interpretation of the changes of the longitudinal relaxation time,
as compared to the mean-field analysis.

\section{Conclusions}
In this paper we have presented a comprehensive description of the critical
dynamics at structurally incommensurate phase transitions. 
Our starting point was the time-dependent, relaxational Ginzburg-Landau model with
$O (2)$ symmetry. To be more general, we discussed the $O
(n)$-symmetric functional. Hence, we were able to study the influence of the
$n-1$ Goldstone modes accurately.
We used the renormalization group theory in order to compute the dynamical
susceptibility below and above the critical temperature $T_I$ 
to one-loop order. Thus we could  venture beyond the usual
mean-field description. 
As we calculated the renormalization factors in
the generalized minimal subtraction scheme,\cite{Ami78,Law81} 
we could deal with the interesting crossover scenarios carefully. \cite{Tae92} 

Our findings were used to interpret experimental data from $NMR$ experiments,
measuring the relaxation rate. The relaxation rate is connected with the
calculated susceptibility via an integral over the wavevector, at fixed
frequency. 

Above the critical temperature $T_I$,
we showed how scaling arguments lead to an identification of the
dynamical critical exponent for the relaxation rate and provide 
a qualitative understanding of its temperature dependence.
Then we described
the crossover from the critical region to a high-temperature region, where
fluctuations should not change the classical critical exponents. 
Excellent agreement for
both the critical exponents resulting from the scaling arguments and the 
description of the crossover regions with the experimental data was found. 
This led
us to the conclusion that the experimental data should probably not 
be interpreted by identifying a critical region of supposed width of $100K$, 
but rather through a crossover between
the non-classical critical exponents and the mean-field exponents, taking place
at a temperature approximately equal to $T_I+5K$. This conjecture
yields a considerably more reasonable width of the critical region.

Below the critical temperature, we analyzed the
dynamical susceptibility calculated to one-loop order in the
renormalization group theory in considerable detail.
The coupling of the $OP$ modes was considered explicitly. 
We thus gained new insight into the influence of 
Goldstone modes on the structure of the susceptibility and its temperature
dependence. 
As a result we found that the relaxation rate of the phason fluctuations 
becomes temperature-dependent. 
This temperature dependence disappears in the two limits when either the
temperature approaches the critical temperature $T_I$, 
or the temperature is very low. 
For the amplitudon modes the influence of the Goldstone modes is more subtle.
We summarized the effect in a wavevector- and frequency-dependent ``mass" and 
showed that this can be interpreted as a bending-down of the temperature-dependent 
relaxation time as compared to a hypothetic situation 
where no Goldstone modes are present. 
All experimental findings are well understood treating the $OP$ modes
beyond their mean-field description. 
As reported from the analysis of ultrasonic attenuation experiments
for Rb$_2$ZnCl$_4$ before, \cite{Sch92,Sch94} no phason gap had to be 
introduced.
Recently however, the direct observation of a ``phason gap" has been reported 
for a molecular compound (BCPS). \cite{Oll98} 
This ``phason gap" was observed in inelastic neutron scattering 
experiments for very high frequencies.
Again, the low frequency dynamics probed by $NMR$ 
did not reveal any gap.\cite{Sou91}
 
Thus, an interesting application 
of the $O(2)$-symmetric model is presented
here, in terms of a crossover description and a discussion of the full $k$ and
$\omega$ dependence of the susceptibility calculated to one-loop order.
We found very good agreement with experimental data.
Besides the precise calculation of critical
exponents as  one strength of the renormalization group theory, also detailed
analysis of crossover scenarios and the effect of the 
anharmonic coupling of modes is possible. 
We want to stress how successfully the results of the renormalization group theory 
can be applied to specific experimental findings. 
In addition, we emphasize that the choice of two fit parameters in the
phase above $T_I$ already essentially determined the curves in the
incommensurate phase.

The theory presented here is formulated in a general way. 
Therefore it could be readily  used to analyze further experiments, 
especially below and near $T_I$.

\acknowledgments
We benefited from discussions with E. Frey, J. Petersson,
and D. Michel. B.A.K. and F.S. acknowledge support from the  
Deutsche Forschungsgemeinschaft under contract No. Schw. 348/6-1,2.
U.C.T. acknowledges support also from the  
Deutsche Forschungsgemeinschaft through a habilitation fellowship
DFG-Gz. Ta 177 / 2-1,2.

\endmcols
\appendix
\section{}
\label{app1}
In this appendix we list analytical results for the two-point vertex functions
and their singularities ($1/\varepsilon$ poles) 
in the generalized minimal subtraction scheme, 
following from the dynamical functional (\ref{dyn_func_harm}).
Below $T_I$, the corresponding zero- and one-loop contributions are stated.
For the explicit calculation of the integrals in
the generalized minimal subtraction scheme we refer to
Ref.  \onlinecite{Fre90}. 
All integrations over internal
frequencies have already been performed by means of the residue theorem.
We restrict ourselves to the three-dimensional case ($\varepsilon =1$). 
\\[0.1cm]
\underline{$T>T_I$:}\\[0.1cm] 
Here only the simplest one-loop graphs enter the diagrammatic expansion.
\begin{align}
\Gamma_{\circ \tilde \psi \psi} ({\vc k}, \omega) 
=& \lambda_\circ \left(k^2 + i \omega/\lambda_\circ + \tau_\circ + 
\frac{n+2}{6} u_\circ \int_{q}^{} \frac{1}{\tau_\circ + q^2}  \right) \nonumber \\
=& \lambda_\circ \left(k^2 + i \omega/\lambda_\circ  
+\tau_\circ \left[ 1- \frac{n+2}{6\varepsilon} u_\circ 
A_d \mu^{-\varepsilon} \frac{\tau_\circ}{(1 +
\tau_\circ/\mu^2)^{\varepsilon/2}} \right] \right) \nonumber \\
=& \lambda_\circ \left(k^2 + i \omega/\lambda_\circ + \tau_\circ/Z_r \right)  
\end{align}
\begin{align}
\Gamma_{\circ\tilde \psi \psi \psi \psi }({\vc k}_i=0, \omega=0) 
=& u_\circ - \frac{n+8}{6} u_\circ \int_{q}^{} \frac{1}{(\tau_\circ + q^2)^2}   
\nonumber \\
= & u_\circ \left[ 1- \frac{n+8}{6\varepsilon} u_\circ 
    A_d \mu^{-\varepsilon} \frac{1}{(1 +
  \tau_\circ/\mu^2)^{\varepsilon/2}} \right] \nonumber \\
= & u_\circ/Z_u 
\end{align}
\underline{$T<T_I$:} \\[0.1cm]
The diagrammatic expressions for the vertex functions are depicted in 
Figs. \ref{fig20} and \ref{fig21}.
\begin{align}
\Gamma_{\circ \ \tilde \pi \pi } ( {\vc k}, \omega) 
=& i \omega + \lambda_\circ \left( k^2 +  \frac{1}{3} u_\circ m_\circ^2
\int_{q}^{} \frac{1}{q^2(m_\circ^2 + q^2)} \right)  
\nonumber \\
&  - \lambda_\circ \frac{1}{3} u_\circ m_\circ^2
\int_{q}^{} \frac{1}{(\frac{{\vc k}}{2} - {\vc q})^2 + m_\circ^2} \cdot 
   \frac{1}{i \omega/\lambda_\circ + (\frac{{\vc k}}{2} + {\vc q})^2 +
   m_\circ^2 + (\frac{{\vc k}}{2} - {\vc q})^2} \nonumber \\
& - \lambda_\circ \frac{1}{3} u_\circ m_\circ^2
\int_{q}^{} \frac{1}{(\frac{{\vc k}}{2} - {\vc q})^2 } \cdot 
   \frac{1}{i \omega/\lambda_\circ + (\frac{{\vc k}}{2} + {\vc q})^2 
    + m_\circ^2+ (\frac{{\vc k}}{2} - {\vc q})^2} \nonumber \\
=& i \omega + \lambda_\circ \left( k^2 + \frac{1}{3} u_\circ A_d m_\circ 
\right)
\nonumber \\
& + \lambda_\circ \frac{1}{3} u_\circ m_\circ^2 \cdot
\frac{A_d}{2}\cdot \frac{1}{k} 
\left( \arcsin \frac{-i\omega/\lambda_\circ + m_\circ^2}{\sqrt{4
m_\circ^2 k^2 + (k^2 +i \omega/\lambda_\circ -  m_\circ^2)^2}} \right.  
\nonumber \\
& +
\left.
\arcsin \frac{k^2 + i\omega/\lambda_\circ-m_\circ^2}{\sqrt{4
m_\circ^2 k^2 + (k^2 +i \omega/\lambda_\circ-m_\circ^2)^2}} \right) 
\nonumber \\
&  + \lambda_\circ \frac{1}{3} u_\circ m_\circ^2 \cdot
\frac{A_d}{2} \cdot \frac{1}{k} 
\left(\frac{\pi}{2} - 
\arcsin \frac{i\omega/\lambda_\circ + m_\circ^2}{k^2 +i
\omega/\lambda_\circ^2 + m_\circ^2} \right) 
\nonumber \\
\end{align}

\begin{align}
\Gamma_{\circ \ \tilde \sigma \sigma } ( {\vc k}, \omega) 
=& i \omega + \lambda_\circ \left( k^2 +  m_\circ^2 \right)  
\nonumber \\
&  - \lambda_\circ \frac{n-1}{3} u_\circ m_\circ^2
\int_{q}^{} \frac{1}{(\frac{{\vc k}}{2} - {\vc q})^2 + m_\circ^2} \cdot 
   \frac{1}{i \omega/\lambda_\circ + (\frac{{\vc k}}{2} + {\vc q})^2 +
   (\frac{{\vc k}}{2} - {\vc q})^2} \nonumber \\
& - \lambda_\circ 3 u_\circ m_\circ^2
\int_{q}^{} \frac{1}{(\frac{{\vc k}}{2} - {\vc q})^2 + m_\circ^2} \cdot 
   \frac{1}{i \omega/\lambda_\circ + m_\circ^2 + (\frac{{\vc k}}{2} + {\vc q})^2 
    + m_\circ^2 + (\frac{{\vc k}}{2} - {\vc q})^2}  \nonumber \\
=& i \omega + \lambda_\circ \left( k^2 +  m_\circ^2 \right)
\nonumber \\
& + \lambda_\circ \frac{n-1}{3} u_\circ m_\circ^2 \cdot
\frac{A_d}{2} \cdot \frac{1}{k} 
\left(\frac{\pi}{2} + 
\arcsin \frac{-i\omega/\lambda_\circ}{k^2 +i \omega/\lambda_\circ} \right) 
\nonumber \\
& + \lambda_\circ 3 u_\circ m_\circ^2 \cdot
\frac{A_d}{2}\cdot \frac{1}{k} 
\left( \arcsin \frac{-i\omega/\lambda_\circ}{\sqrt{4
m_\circ^2 k^2 + (k^2 +i \omega/\lambda_\circ)^2}} \nonumber \right. \\
& \left. +
\arcsin \frac{k^2 + i\omega/\lambda_\circ}{\sqrt{4
m_\circ^2 k^2 + (k^2 +i \omega/\lambda_\circ)^2}} \right) 
\nonumber  \\
\end{align}

\beginmcols

\begin{table}
\begin{tabular}{c|c|c}
$\nu = 0.6695$ & $\eta = 0.033$ & $z = 2.024$  \\ 
\end{tabular}
\caption{Critical exponents in the $O(2)$ symmetric $\phi^4$ field theory
in three dimensions. \cite{Hoh77,Zin93}}
\label{tab1}
\end{table}

\begin{table}
\begin{tabular}{l|l|l|l}
$\omega$ &  $k$ & $\chi_\parallel^{-1}$ & $ \Im \chi_\parallel^{-1} $\\
\hline
$\omega \rightarrow 0$ & $k \rightarrow 0$ & $\propto k $
& $ \rightarrow 0 $ \\[0.3cm]
\hline
$\omega \neq 0$ & $k \rightarrow \infty$ & $\propto k^2 $
& $\propto k^4 $ 
\\[0.3cm]
& $\sqrt{\omega/\lambda} < k < m(1)$ 
& $- i \omega/\lambda + 
\frac{m(1)^2 \cdot k}{1+ a f(k)}$& $\propto m(1)^4 k^3$\\[0.3cm]
& $k \leq \sqrt{\omega/\lambda} $
&  $- i \omega/\lambda + 
\frac{m(1)^2 \cdot k}{1+ a f(k)}$
& $\propto m(1)^{-2}$ \\[0.3cm]
& $k \rightarrow 0$ & $ - i \omega/\lambda + \frac{2 k \cdot 
m(1)^2 }{a} $ 
& $\omega/\lambda $
\end{tabular}
\caption{Different scaling laws for the longitudinal susceptibility. 
The functions $f(k)$ denote regular functions.}
\label{tab2}
\end{table}

\begin{figure}
\begin{picture}(11,7)
\put(120,6.5){\makebox(0,0){$\times \qquad A=-\frac{n-1}{6}
u_\circ \int_k^{} \frac{1}{k^2} - \frac{1}{2} u_\circ \int_k^{}
\frac{1}{m_\circ^2 + k^2}$}} 
\end{picture}
\caption{Counterterm  below $T_I$.}
\label{fig8}
\end{figure}

\begin{figure}
\epsfig{file=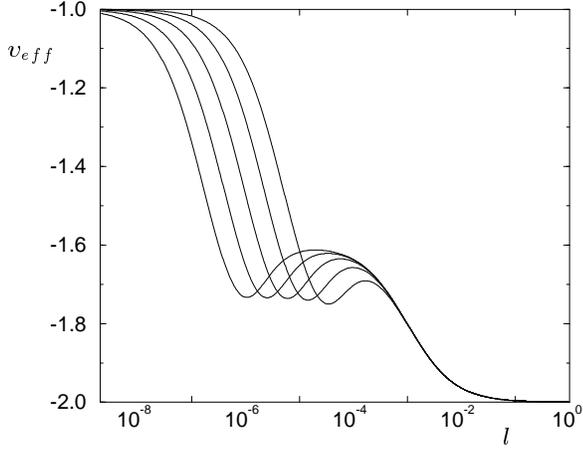, width=8.0cm}
\caption{The effective exponent 
$\upsilon_{eff}=\frac{\partial \log m(l)}{\partial \log l}$ of
the flow of the renormalized mass parameter 
$m(l)$ vs. $l$ for seven different $m(1) \ll 1 $ below $T_I$ 
[$u(1)=10^{-2} \cdot u_H^*, n=2, \varepsilon=1$]}
\label{fig10}
\end{figure}

\begin{figure}
\epsfig{file=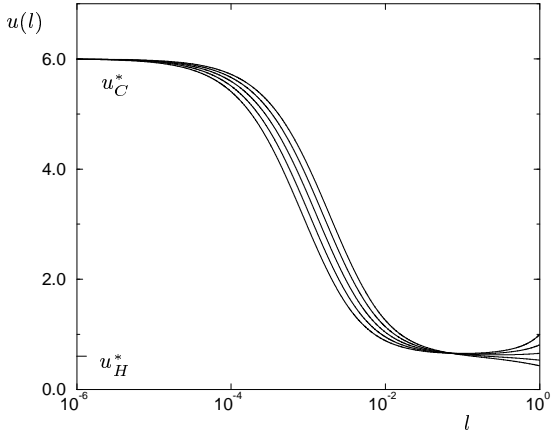, width=8.5cm}
\caption{Flow of the coupling constant
$u(l)$ vs. $l$ for five different $u(1)$ below $T_I$ 
[$m(1)=0.01, n=2, \varepsilon=1$]. The coexistence and the Heisenberg fixed
point are marked on the $u$ axis.}
\label{fig11}
\end{figure}

\begin{figure}
\epsfig{file=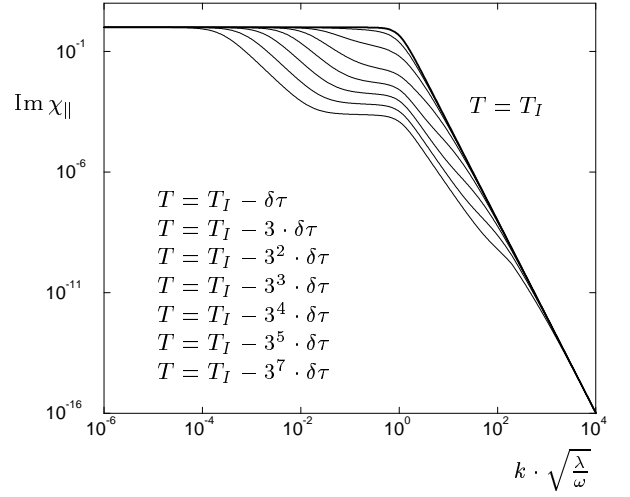, width=8.5cm}
\caption{The longitudinal susceptibility for the critical temperature $T = T_I$
and seven other temperatures plotted
vs. $k \cdot \sqrt{\frac{\lambda}{\omega}}$. The scaling regions listed in
Table \ref{tab2} are valid for temperatures not too close to $T_I$.}
\label{fig13}
\end{figure}

\begin{figure}
\epsfig{file=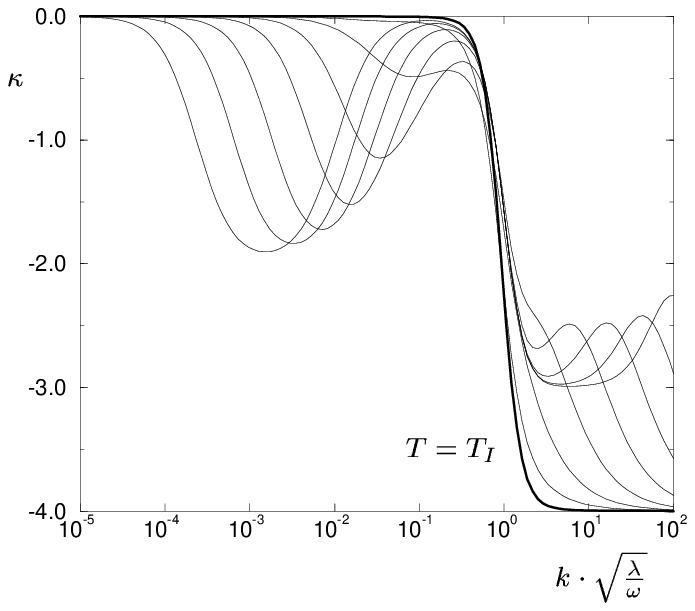, width=8.5cm}
\caption{The effective exponent $\kappa$ of $\Im \chi_\parallel$ with respect
to $k$ plotted vs. $k \cdot \sqrt{\frac{\lambda}{\omega}}$ for the same
temperatures as in Fig. \ref{fig13}.}
\label{fig14}
\end{figure}

\begin{figure}
\epsfig{file=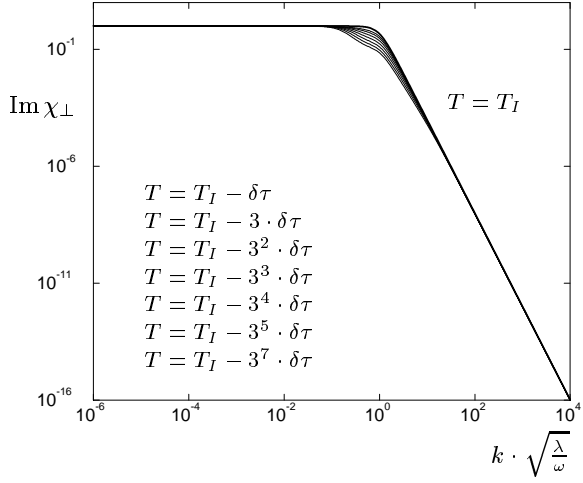, width=8.5cm}
\caption{The transverse susceptibility for the critical temperature $T = T_I$
and seven other temperatures plotted vs. $k \cdot
\sqrt{\frac{\lambda}{\omega}}$. The temperature dependence is
only present in an interim region.}
\label{fig15}
\end{figure}

\begin{figure}
\epsfig{file=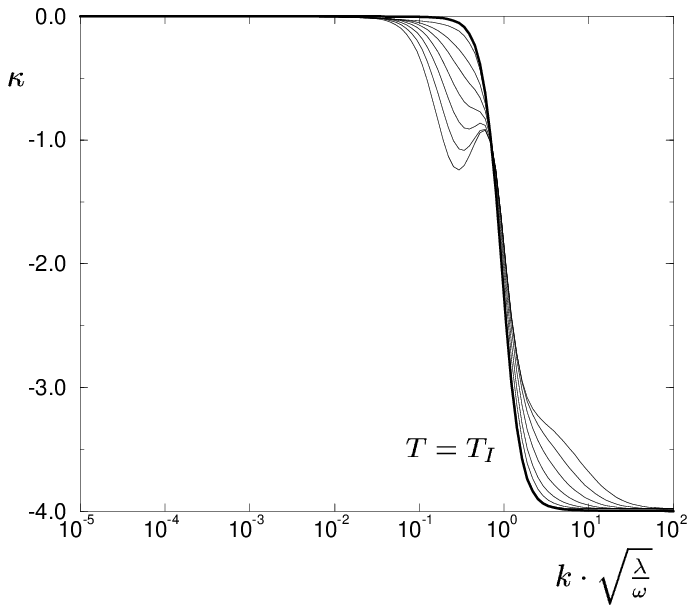, width=8.5cm}
\caption{The effective exponent $\kappa$ of $\Im \chi_\perp$ with respect
to $k$ plotted vs. $k \cdot \sqrt{\frac{\lambda}{\omega}}$ for the same
temperatures as in Fig. \ref{fig15}.}
\label{fig16}
\end{figure}

\begin{figure}
\epsfig{file=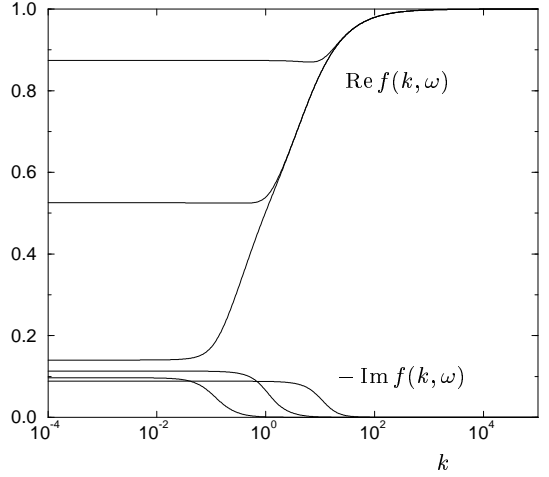, width=8.5cm}
\caption{Real and imaginary part of the effective "mass" (self energy) $f$
and its $k$ dependence for $\tilde \omega =0.01, 1,$ and $100$.}
\label{fig17}
\end{figure}

\begin{figure}
\epsfig{file=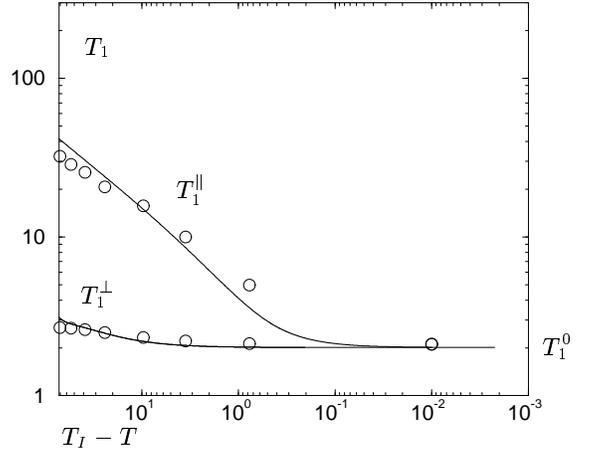, width=8.cm}
\caption{Relaxation time $T_1^\perp$ (phason dominated) and $T_1^\parallel$ 
(amplitudon dominated) vs. reduced temperature. The $NMR$ experimental data,
\cite{Mis97} again indicated by circles, are compared with the theoretical results,
represented by the solid lines. As in the high-temperature phase 
a crossover to the mean-field regime starting
at $\Delta T \approx 5 K$ is considered.}
\label{fig18}
\end{figure}

\endmcols

\begin{figure}
\centering
\epsfig{file=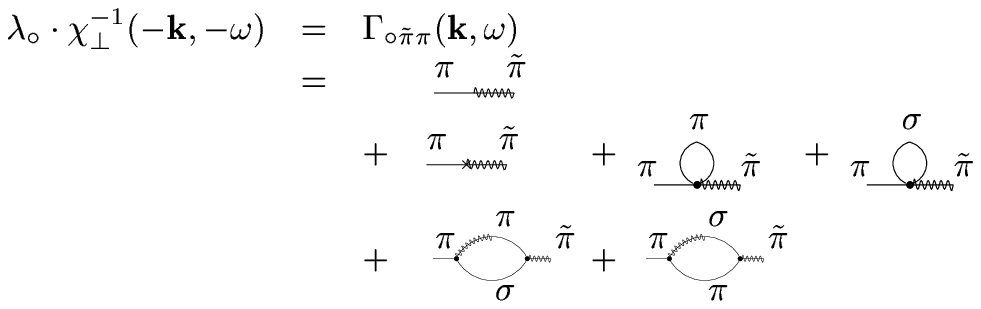, width=10.5cm}
\caption{Transverse susceptibility and its diagrammatic representation.}
\label{fig20}
\end{figure}

\begin{figure}
\centering
\epsfig{file=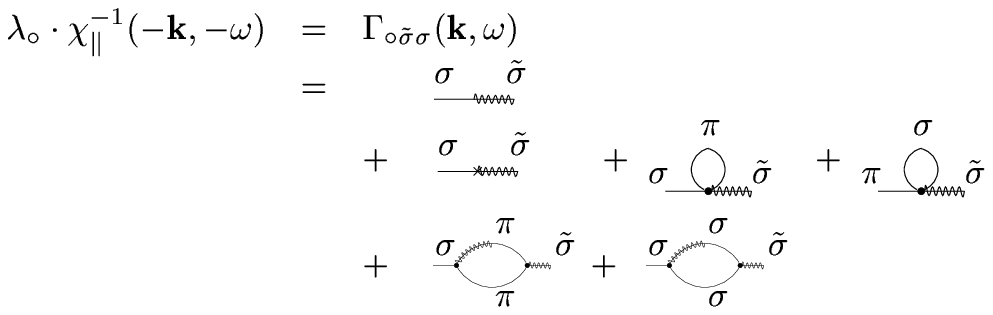, width=10.5cm}
\caption{Longitudinal susceptibility and its diagrammatic representation.}
\label{fig21}
\end{figure}

\end{document}